\documentclass[article]{aa}

\usepackage{graphicx}
\usepackage{amsmath,systeme}

\usepackage{txfonts}
\usepackage{hyperref}
\usepackage{natbib}
\bibpunct{(}{)}{;}{a}{}{,} 
\usepackage{color}
\usepackage{xspace}

\usepackage{academicons}
\usepackage{xcolor}
\newcommand{\orcid}[1]{}

\usepackage{listings}

\definecolor{orange}{rgb}{1.0, 0.4980392156862745, 0.054901960784313725}
\definecolor{codegreen}{rgb}{0,0.6,0}
\definecolor{codegray}{rgb}{0.5,0.5,0.5}
\definecolor{codepurple}{rgb}{0.58,0,0.82}
\definecolor{backcolour}{rgb}{0.95,0.95,0.92}

\lstdefinestyle{mystyle}{
    backgroundcolor=\color{backcolour},
    commentstyle=\color{codegreen},
    keywordstyle=\color{magenta},
    numberstyle=\tiny\color{codegray},
    stringstyle=\color{codepurple},
    basicstyle=\ttfamily\footnotesize,
    breakatwhitespace=false,
    breaklines=true,
    captionpos=b,
    keepspaces=true,
    numbers=left,
    numbersep=5pt,
    showspaces=false,
    showstringspaces=false,
    showtabs=false,
    tabsize=2
}

\newcommand{\Gaia}{{\it Gaia}\xspace}

\newcommand{\mpia}{Max-Planck-Institut f\"ur Astronomie, K\"onigstuhl 17, D-69117 Heidelberg, Germany}
\newcommand{\leiden}{Leiden Observatory, Leiden University, Niels Bohrweg 2, 2333 CA Leiden, The Netherlands}
\newcommand{\cambridge}{Institute of Astronomy, University of Cambridge, Madingley Road, Cambridge CB3 0HA, United Kingdom}

\newcommand{\torino}{INAF - Osservatorio Astrofisico di Torino, Strada Osservatorio 20, Pino Torinese 10025 Torino, Italy}

\newcommand{\Monash}{School of Physics \& Astronomy, Monash University, Clayton 3800, Victoria, Australia}
\newcommand{\AstroTD}{Centre of Excellence for Astrophysics in Three Dimensions (ASTRO-3D), Melbourne, Victoria, Australia}
\begin{document}

   \title{Uniting \Gaia and APOGEE to unveil the cosmic chemistry of the Milky Way disc}

    \titlerunning{Combining \Gaia and APOGEE}
    \authorrunning{Cantat-Gaudin et al.}

   \author{Tristan Cantat-Gaudin\inst{1} \orcid{0000-0001-8726-2588}
          \and
          Morgan Fouesneau\inst{1}  \orcid{0000-0001-9256-5516}
          \and
          Hans-Walter Rix\inst{1} \orcid{0000-0001-5996-8700}
          \and
          Anthony G. A. Brown\inst{2} \orcid{0000-0002-7419-9679}
            \and
            Ronald Drimmel\inst{3} \orcid{0000-0002-1777-5502}
          \and
            Alfred Castro-Ginard\inst{2} \orcid{0000-0002-9419-3725}
            \and
            Shourya Khanna\inst{3} \orcid{0000-0002-2604-4277}
            \and
            Vasily Belokurov\inst{6} \orcid{0000-0002-0038-9584}
            \and
            Andrew R. Casey\inst{7,8} \orcid{0000-0003-0174-0564}
          }

   \institute{
        \mpia  \\ \email{cantat@mpia.de}
         \and \leiden \and \torino \and \cambridge \and \Monash \and \AstroTD  
             }

   \date{}

 \abstract{The spatial distribution of Galactic stars with different chemical abundances encodes information on the processes that drove the formation and evolution of the Milky Way. Survey selection functions are indispensable for analysing astronomical catalogues produced by large-scale surveys. The use of these selection functions in data modelling is more complex when data from different surveys are to be modelled simultaneously. We introduce a procedure for constructing the selection function of a sample of red clump stars that have parallaxes and elemental abundances from the \Gaia mission. We separately constructed the selection function of the APOGEE DR17 red clump stars, which depends on very different observables and has a very different spatial coverage. We combined the two surveys and accounted for their joint selection function to provide strong constraints on the radial and vertical density distribution of mono-abundance populations, with \Gaia offering a dense coverage of the solar neighbourhood, while APOGEE reaches larger distances near the Galactic plane. We confirm that the radial density profile steepens with increasing metallicity. The combined sample also indicates a metallicity-dependent flaring of the $\alpha$-poor disc. We provide the code for constructing the \Gaia selection function we used in this study through the \texttt{GaiaUnlimited} Python package.}




   \maketitle
%

\section{Introduction}\label{sec:intro}

The spatial and chemical distribution of stars in the stellar disc of the Milky Way results from the complex and intricate mechanisms that formed and shaped the evolution of our Galaxy. By exploring the chemical diversity across different stellar populations, we can glean valuable insights into the formation, enrichment, and evolution of galaxies, shedding light on fundamental astrophysical processes \citep{Gilmore89,Bovy12MAPs,RixAndBovy13,Hayden15,Bovy16population}.  

Investigation of the detailed chemical composition of stars throughout the Milky Way enables us to discern signatures of subsequent stellar generations and their respective nucleosynthetic contributions, and it provides insight into the formation history and enrichment mechanisms of various stellar populations within our Galaxy \citep[e.g.][]{Freeman02}. Ultimately, this knowledge allows us to build a timeline of galactic evolution, and to uncover the role played by different progenitors, such as massive stars, novae, and supernovae, in shaping the chemical landscape. By comparing observed chemical abundances with predictions from stellar evolution and galactic chemical evolution models, we can refine our understanding of the physical processes involved \citep[e.g.][]{Bensby14, Martig16}. This iterative process allows us to improve our models, leading to more realistic simulations of stellar populations, galactic structures, and the chemical evolution of galaxies.

Many of these insights arise from asking how many stars of which abundance (and possibly which age) are observed in different parts of the Milky Way. This information is provided by large spectroscopic surveys. To analyse these surveys, their survey selection function is pivotal to understand which stars at which position in the Galaxy were able to be included in any survey under consideration \citep[e.g.][]{Rix21}. These functions consider various observational constraints, such as magnitude limits, sky coverage, signal-to-noise ratios, and target selection algorithms. Neglecting or misinterpreting selection functions can lead to distorted conclusions about the chemical distribution of stars and their implications for galactic evolution. The present paper is part of a series in the \emph{GaiaUnlimited} project\footnote{\url{http://gaia-unlimited.org}}, which aims to provide the astronomical community with tools and methods to account for the selection function of various subsets of the \Gaia data \citep[see also][]{Rix21, CantatGaudin23, CastroGinard23_subsamples}.

In this study, we focus on red clump stars, which are low-mass stars in the core helium-burning stage that are located in a distinct region of the Hertzsprung-Russell diagram and exhibit remarkable homogeneity. They are therefore excellent standard candles  \citep[][]{Girardi16, Hawkins17, Hall19, Chan20}, and their chemical abundances allow us to investigate the radial abundance gradients in the Milky Way disc. In particular, examining the spatial variations in their chemical composition in different galactic regions provides insights into the kinematic properties of stellar populations, the role of migration, and the formation history of the galactic disc. 
We combine precise chemical abundances and parallaxes from \Gaia DR3 \citep[][]{gaiaMission,GaiaDR3} with abundances from the Sloan Digital Sky Survey (SDSS) Apache Point Observatory Galactic Evolution Experiment (APOGEE) seventeenth data release \citep[DR17;][]{Majewski17,Abdurrouf22}. The combination of these two surveys offers a wide sampling of the Milky Way disc, with \Gaia providing a dense coverage of the solar neighbourhood, while APOGEE offers a sparser but deeper coverage of the inner and outer disc. The aim of our paper is two-fold: We provide an example of how very different survey selection functions of surveys can be combined so that they can be modelled jointly. We also refine our understanding of the radial structure of mono-abundance populations \citep{Bovy12MAPs} in the Galactic disc.

This paper is structured as follows. Sect.~\ref{sec:apogeeRC} introduces the sample of APOGEE red clump stars and its selection function. Sect.~\ref{sec:gaiadata} builds a sample of \Gaia DR3 red clump stars with precise elemental abundances from the \Gaia GSP-Spec pipeline, and Sect.~\ref{sec:gaia_sf} details the construction of the corresponding selection function. In Sect.~\ref{sec:density}, we then explore the density distribution of mono-abundance populations, and we discuss our results in Section~\ref{sec:discussion}. Sect.~\ref{sec:conclusion} closes with concluding remarks.

%
%
\section{Data sets and their selection functions}\label{Data_and_SF}

In this section we introduce the two data sets that we used to determine the radial disc profiles for stellar mono-abundance populations in the Galactic disc. We start with the SDSS/APOGEE sample, and then we turn to the \Gaia sample, for which the selection function has not been worked out before. For Gaia, determining the selection function, that is, the probability of any actual or counterfactual star to be in the sample, is a multistep process.

\subsection{The SDSS/APOGEE Data} \label{sec:apogeeRC}

\begin{figure*}
\includegraphics[width=0.99\textwidth]{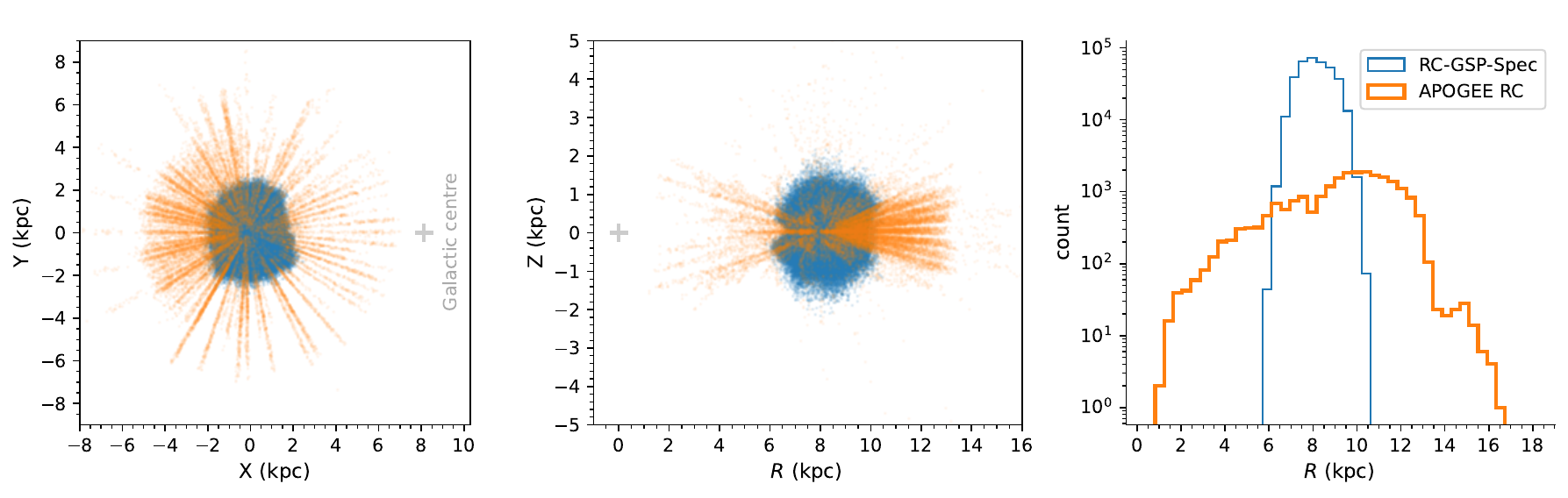} 
\caption{ Spatial distribution of the SDSS/APOGEE red clump stars and \Gaia RC-GSP-Spec stars (see Sect.~\ref{sec:gaiarc}). Left: location projected on the Galactic plane. Middle: Galactic altitude $Z$ vs. Galactocentric distance $R$. Right: Distribution of Galactocentric distance $R$.  \label{fig:blue_orange_panels}
}
\end{figure*}

We employed the data from the 17th data release \citep[hereafter DR17]{Abdurrouf22} of the SDSS survey, focussing on the APOGEE subset \citep[][]{Majewski17}, which collected high-resolution near-infrared spectroscopic data for cool stars (in particular, red giants). All data used in this study were accessed through the SDSS portal\footnote{\url{https://www.sdss4.org/dr17/irspec/spectro_data/}}. The identification of red clump stars in SDSS/APOGEE is described in \citet{Bovy14}. Fig.~\ref{fig:blue_orange_panels} shows their spatial and radial distributions.

\subsubsection{APOGEE Main Red Star Sample selection function} \label{sec:apogee_sf}

The complex selection function for the APOGEE targets in SDSS is described in \citet{Zasowski13} and \citet{Zasowski17}. The selection function of the entire survey is not tractable in principle, but
we take advantage of the fact that the APOGEE Main Red Star Sample was selected by randomly choosing targets from the 2MASS catalogue \citep[Two Micron All Sky Survey;][]{Skrutskie16} in specific regions of the $( (J-K_s)_0, H)$ colour-magnitude diagram, corresponding to the photometric space of the reddening-corrected $(J-K_s)$ colour and the observed (uncorrected) $H$ magnitude. This enables us to construct its selection function. 

The Main Red Sample stars can be identified in the APOGEE DR17 table with the \texttt{EXTRATARG} flag set to 0, \texttt{PROGRAMNAME} to \texttt{bulge}, \texttt{disk}, \texttt{disk1}, \texttt{disk2} or \texttt{apogee}, and a \texttt{FIELD} name following the pattern \texttt{L+B} or \texttt{L-B} (e.g. \texttt{120+12}), which indicates that it is part of the semi-regular grid footprint. About half of the 730,000 DR17 sources are part of pointings like this.
To construct the selection function of this sample, we followed the general approach introduced in Section~4 of \citet{Bovy14} \citep[see also][]{Bovy16population, Frankel19growth}, which we summarise here.

\begin{figure*}
\includegraphics[width=0.99\textwidth]{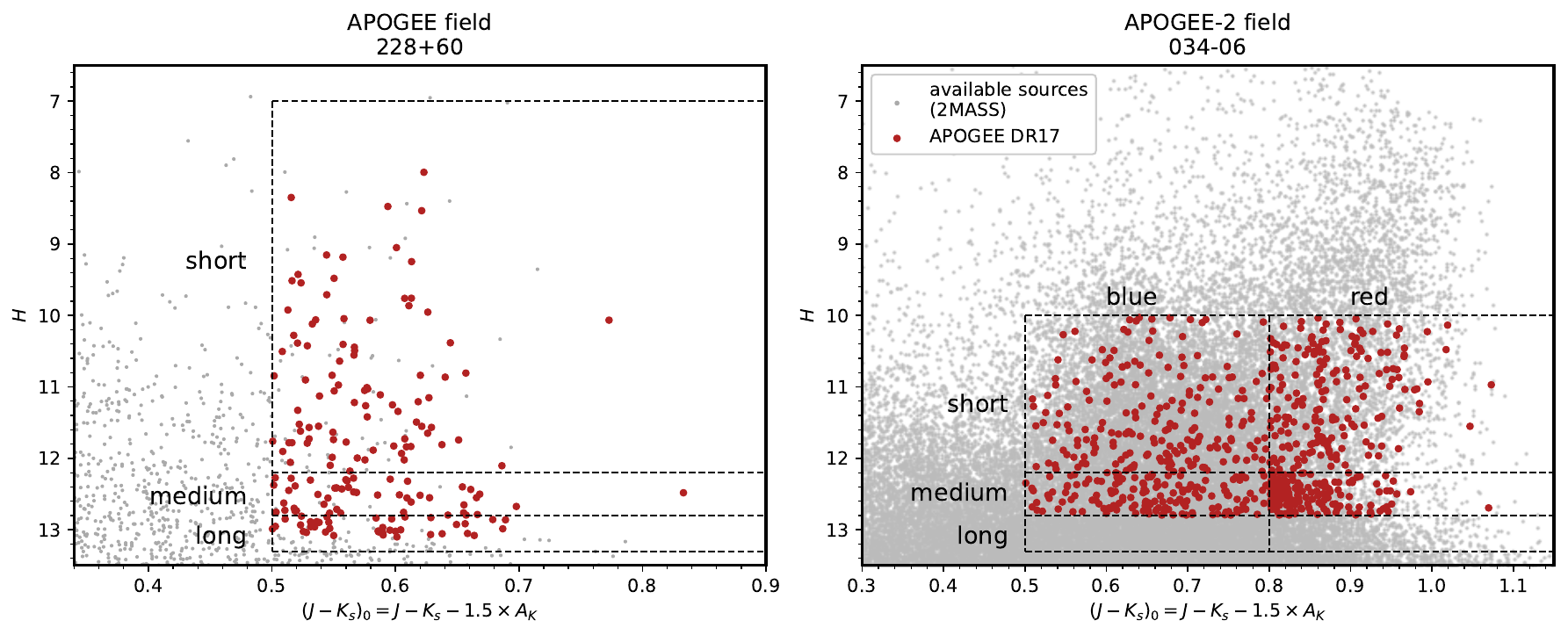} 
\caption{ Binning in intrinsic colour-magnitude space used by APOGEE (left) and APOGEE-2 (right) to select their Main Red Star Sample targets from the 2MASS catalogue. The exact magnitude limits between the three cohorts (short, medium, and long) depend on the field and are provided by the APOGEE collaboration in auxiliary files. The selection is computed as the ratio of successfully observed targets to available sources in each of the three (in APOGEE) or six (in APOGEE-2) regions. \label{fig:apogee_cohorts}}
\end{figure*}

In each field observed by APOGEE, the survey selected targets from three different magnitude ranges, which are referred to as short cohort, medium cohort, and long cohort. These cohorts typically correspond to $H$-magnitude ranges [7,12.2] ([10,12.2] in APOGEE-2), [12.2,12.8], and [12.8,13.3], respectively. The details of the cohort definition of each field are listed in the survey \texttt{allDesign} table\footnote{\url{https://data.sdss.org/sas/dr17/apogee/target}}. The design table also specifies the colour range of each cohort. For APOGEE-1 observations, it is typically a single $(J-K_s)_0 > 0.5 $ colour limit. For APOGEE-2, each cohort exists in two different colour ranges, typically, $0.5 < (J-K_s)_0 < 0.8$ and $(J-K_s)_0 > 0.8 $, which are referred to as blue and red (see Fig.~\ref{fig:apogee_cohorts}).

For each design, the APOGEE survey selected a random subset of $\sim$150 targets from the 2MASS data. The exact number depends on whether other science programmes have targets of interest in a given field of view, but it is ultimately limited by the number of available spectrograph fibers (230 in APOGEE, and 250 in APOGEE-2). In order to collect larger numbers of spectra, many fields have been observed multiple times with different designs (and a different corresponding target list). The selection function for these fields can only be constructed when every associated design is constructed with the same $H$ and $(J-K_s)_0$ limits. In some rare cases, the radius of the field or the central coordinates of the pointing were also changed between designs, which means that the selection function is intractable for these fields as well.

The APOGEE survey conveniently shares\footnote{In directories \texttt{apogeeObject} and \texttt{apogee2Object} at \url{https://data.sdss.org/sas/dr17/apogee/target}} the 2MASS data used for the target selection in each field, along with the extinction factor\footnote{ The values of $A_K$ were obtained from multiple sources and are discussed in Sect.~4.3.1 of \citet{Zasowski13}} $A_K$ used to obtain $(J-K_s)_0 = J - K - 1.5 \cdot A_K$ \citep{Indebetouw05}. To obtain the selection fraction in each cohort, we simply computed the ratio of successfully observed stars to those available in 2MASS in the corresponding region of the intrinsic colour-magnitude space. We also accounted for the fact that in some cases (especially in the long cohorts of bulge fields), a fraction of the 2MASS stars has no associated value of $A_K$ and therefore could not be targeted by APOGEE.

We find a very good agreement between our results and those published by \citet{Frankel19growth} for the short cohorts of APOGEE DR14 (except that DR17 covers more fields, and some of them are more complete). 
We make the final table available through the \texttt{gaiaunlimited}\footnote{\url{https://github.com/gaia-unlimited/gaiaunlimited}} Python package, along with a module to query it at any location, magnitude, and colour.

\subsubsection{Effective selection function of the APOGEE red clump sample} \label{sec:apogee_effective_sf}

Section~\ref{sec:apogee_sf} expresses the APOGEE selection function in terms of survey observables $(\ell,b,H,(J-K)_0)$. For a stellar population of a given absolute magnitude, we can construct the effective selection function in physical coordinates $(\ell,b,d)$ where $d$ is the distance from the Sun.

\begin{figure}
\includegraphics[width=0.99\columnwidth]{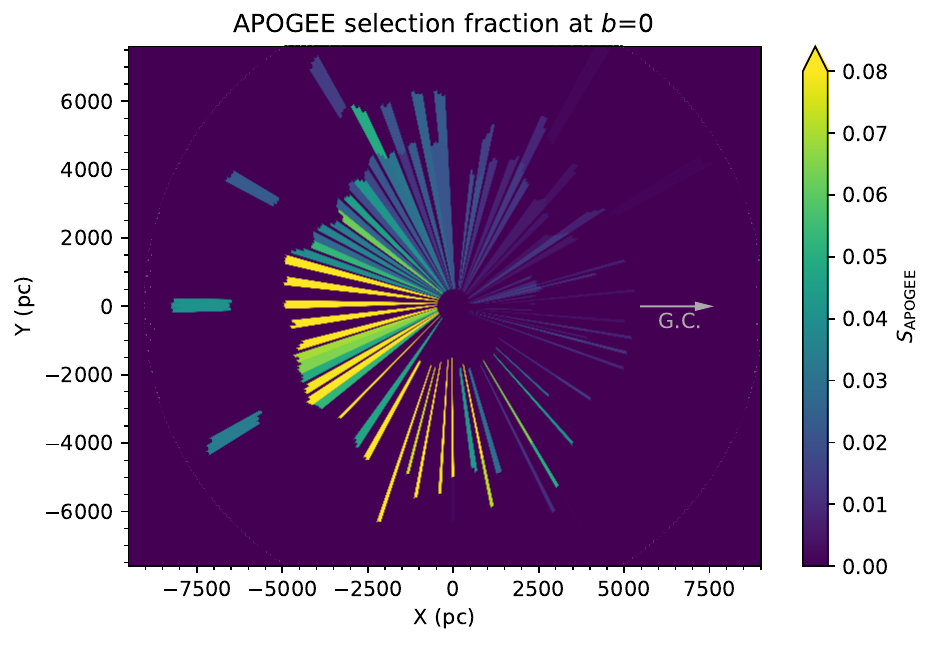} 
\caption{ Effective selection function for the APOGEE DR17 red clump stars shown in the Galactic $(X,Y)$ plane at $b$=0. \label{fig:apogee_sf_xy} Each pencil beam corresponds to a different APOGEE pointing. The difference in bright magnitude limit between APOGEE-1 ($H$=7\,mag) and APOGEE-2 ($H$=10\,mag) is visible in the fourth quadrant (lower left corner).}

\end{figure}

We assumed that red clump stars have an absolute magnitude $H_0$=-1.5\,mag and intrinsic colour $(J-K)_0$=0.60\,mag, placing them in the bluer bin of the APOGEE-2 cohorts. We computed the expected apparent magnitude $H$ for a red clump star at any location $(\ell,b)$ and distance using the \citet{Lallement22} extinction map and assuming $A_H = 0.1727 \times A_0 + 0.0201$\footnote{We calibrated this relation on the sample of APOGEE red clump stars, for which $H_0$ is provided. }. Fig.~\ref{fig:apogee_sf_xy} shows a top-down view of the complex, discontinuous pattern of the effective selection function at $b$=0. The difference in bright magnitude limit between APOGEE-1 ($H$=7\,mag) and APOGEE-2 ($H$=10\,mag) is especially visible in the fourth quadrant (lower right).

%
%
\subsection{\Gaia red clump stars and chemical abundances} \label{sec:gaiadata}

In this section, we first build a sample of bona fide red clump stars, and we then select those with the most reliable metallicities and $\alpha$ abundances and calibrate these abundances to the APOGEE abundance scale.

\subsubsection{\Gaia red clump stars} \label{sec:gaiarc}

We selected red clump stars from the \Gaia DR3 catalogue from their location in a colour-absolute magnitude diagram, that is, the colours and magnitude were not corrected for extinction but corrected for distance according to their parallaxes \citep[using the prescriptions of][]{Lindegren21parallaxbias}. This study later focuses on stars with \Gaia DR3 chemical abundances (see Sect.~\ref{sec:gspspec}), which are all brighter than $G$=14\,mag and closer than 3\,kpc (80\% are closer than 1.5\,kpc). The mean parallax uncertainty at this magnitude is 0.016\,mas, making the difference between the true and the estimated absolute magnitude negligible. This allows us to neglect the impact of parallax uncertainties on the selection of red clump stars.

\begin{figure}
\includegraphics[width=0.99\columnwidth]{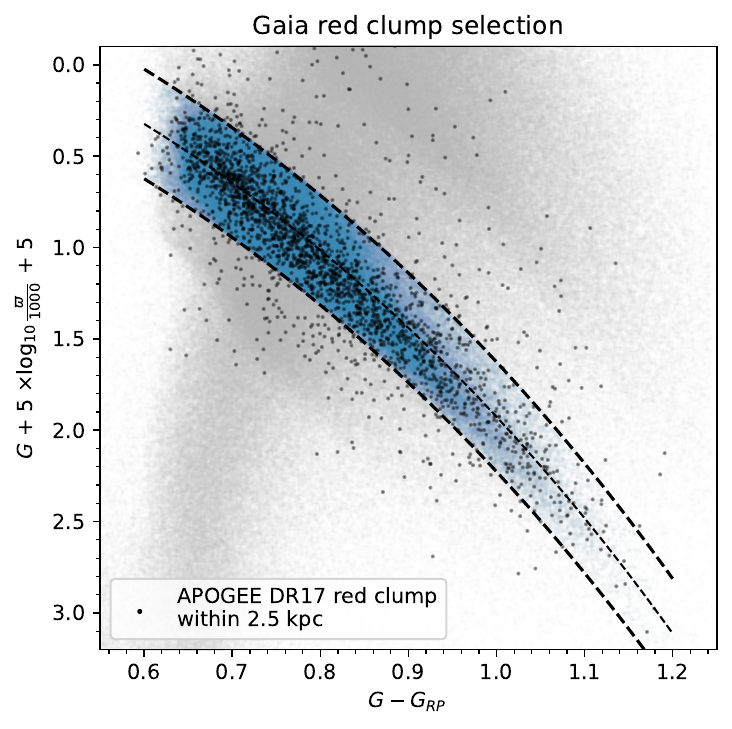} 
\caption{Definition of the \Gaia red clump star sample (see Sect.~\ref{sec:gaiarc}). The shaded area between the dashed lines represents our selection from the \Gaia DR3 photometry described in \ref{sec:gaiarc}, based upon the APOGEE DR17 red clump candidates indicated by the black dots. The grey background indicates the density of all sources in the ``gradient sample'' of \citet{GaiaDR3chemcartoPVP} for reference. \label{fig:gaia_rc_selection}}  
\end{figure}

To account for the effect of extinction on the magnitude and $(G-G_{RP})$ colour of the RC stars, we first visualised the colour-magnitude distribution of stars in common between \Gaia and the APOGEE DR17 catalogue of red clump stars (introduced in Sect.~\ref{sec:apogeeRC}). We fit a fifth-order polynomial to the locus traced by the APOGEE RC stars and considered stellar sources within $\pm$0.3\,mag\footnote{This selection roughly corresponds to twice the dispersion of 0.17\,mag in $M_G$ reported by \citet{Hawkins17}. } of this trace that satisfied $0.6 < (G-G_{RP}) < 1.2 $ to be bona fide RC stars (Fig.~\ref{fig:gaia_rc_selection}). Assuming Gaussian parallax errors, our selection retains over 99.9\% of the red clump stars with $G<14$.

\subsubsection{The RC from the GSP-Spec abundance sample} \label{sec:gspspec}

The \texttt{gaiadr3.astrophysical\_parameters} table from the \Gaia archive\footnote{\url{https://gea.esac.esa.int/archive/}} provides stellar parameters and elemental abundances from the General Stellar Parametriser from spectroscopy \citep[hereafter GSP-Spec;][]{RecioBlanco23gspspec} analysis of five million sources observed with the Radial Velocity Spectrometer \citep[RVS;][]{Cropper18}. For \Gaia~DR3, the GSP-Spec pipeline characterised sources whose combined spectrum has a signal-to-noise ratio greater than 20, effectively favouring brighter sources and sources that received more visits.
From this sample, we further selected the gradient analysis sample, which is defined in the chemical cartography study of \citet{GaiaDR3chemcartoPVP}. By construction, this sub-sample supports the analysis of the relations between chemistry and Milky Way structure, stellar kinematics, and orbital parameters. In particular, this sample corresponds to a subset of stars with good GSP-Spec quality flags \citep[\texttt{flags\_gspspec}\footnote{\url{https://gea.esac.esa.int/archive/documentation/GDR3/Gaia_archive/chap_datamodel/sec_dm_astrophysical_parameter_tables/ssec_dm_astrophysical_parameters.html}};][]{RecioBlanco23gspspec} that satisfy the astrometric quality criterion \texttt{ruwe}<1.4 \citep{Lindegren21astrosolution}.  
Listing~3 in Appendix~B of \citet{GaiaDR3chemcartoPVP} gives the ADQL query to retrieve this sample.

We further selected the stars with total abundance uncertainties in metallicity and an $\alpha$-element abundance smaller than 0.1\,dex to be able to map the spatial distribution of mono-abundance populations,

\begin{equation}
\sqrt{  e_{[Fe/H]}^2 + e_{[\alpha/Fe]}^2  } < 0.1 {\rm dex},
\label{total_uncert_req}
\end{equation}
\noindent where
\begin{equation}
  e_{[Fe/H]} =\ \texttt{mh\_gspspec\_upper}-\texttt{mh\_gspspec\_lower}
\end{equation}
\noindent and
\begin{equation}  
\begin{split}
  e_{[\alpha/Fe]} = & \ \texttt{alphafe\_gspspec\_upper} \\ & \hspace{30pt} -\texttt{alphafe\_gspspec\_lower}
\end{split} 
\end{equation} 

\noindent are the widths of the 68\% confidence intervals on iron abundance and $\alpha$ abundance listed in the \Gaia DR3 astrophysical parameter table, respectively.


In the rest of this paper, we refer to this sample of 356,755 stars as the RC-GSP-Spec sample. Fig.~\ref{fig:blue_orange_panels} shows the spatial distribution of these stars in blue.

The selection based on the abundance precision indirectly corresponds to a selection in signal-to-noise ratio in the RVS spectra, which to first order depends on the apparent $G$ magnitude and on the location on the sky (the RVS spectra of stars in regions that are visited more often by \Gaia are combined from a larger number of exposures). Because the spectra of metal-rich stars have more prominent absorption lines, their abundances tend to be more precise, while spectra with weaker lines require a high signal-to-noise ratio to reach the same abundance precision. Sect.~\ref{sec:SF_met} shows that the selection has a more stringent effect on metal-poor stars, and we account for the varying depth of the selection function with metallicity.

\subsubsection{Calibrating the GSP-Spec abundances on the APOGEE scale} \label{sec:calibration}

\begin{figure*}
\includegraphics[width=0.99\textwidth]{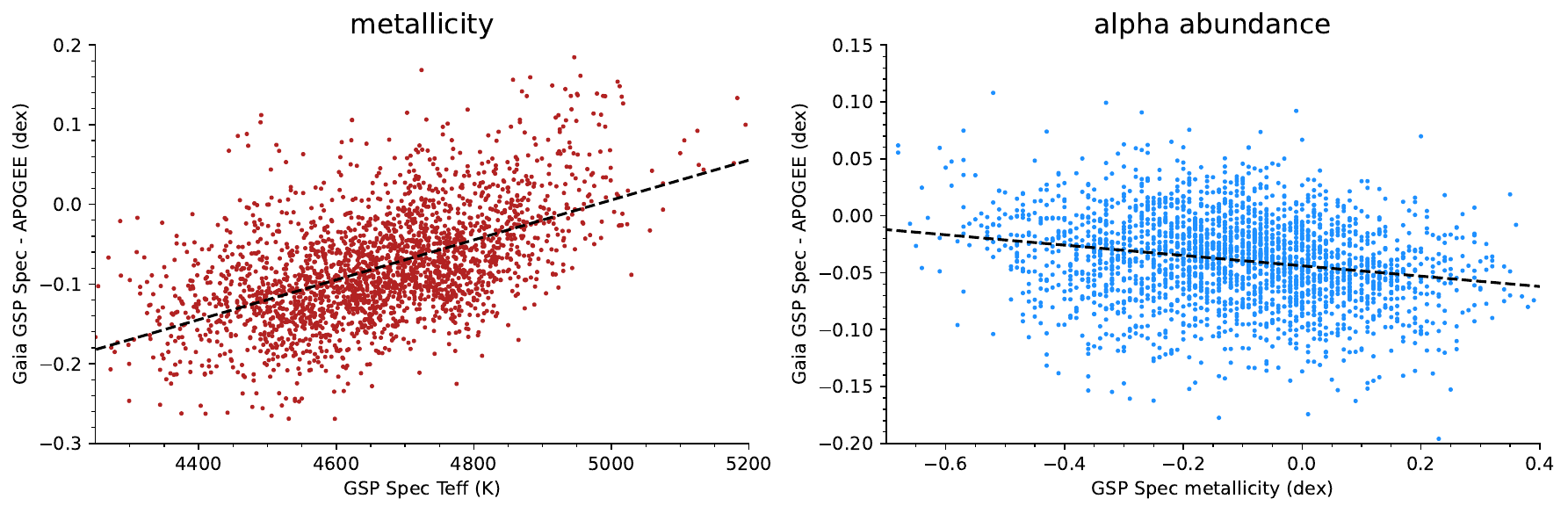} 
\caption{ 
Abundance-scaling relations of \Gaia DR3 GSP-Spec and APOGEE DR17 ASCAP abundances for a sample of 2\,450 red clump stars in common. 
Left: Offset between \Gaia DR3 \texttt{mh\_gspspec} metallicities and APOGEE DR17 ASCAP [M/H]. Right: Offset between [$\alpha$/Fe] (from GSP-Spec) and [Ca/Fe] (from APOGEE). In both panels, the dashed line indicates the correction we applied to calibrate \Gaia on the APOGEE scale, given in equations \ref{eq:correctionMH} and \ref{eq:correctionALPHA}. \label{fig:calibration_abundances}}
\end{figure*}

We find 2\,450 red clump stars in common in our sample and APOGEE DR17. 
We compared the metallicities reported by both catalogues, and we find that
the GSP-Spec metallicities are 0.08\,dex lower than the APOGEE values (with a dispersion of 0.07\,dex) with a strong temperature dependence (left panel of Fig.~\ref{fig:calibration_abundances}), which is consistent with the biases reported by \citet{RecioBlanco23gspspec}. We therefore calibrated the \Gaia metallicity scale on APOGEE by fitting a linear relation between their offset and the GSP-Spec effective temperature, 

\begin{equation} \label{eq:correctionMH}
    \Delta [M/H] = 2.988 \times 10^{-4} \times \left( T_{\mathrm{eff}} - 4700 \right) - 7.264 \times 10^{-2} 
\end{equation}
where $\Delta [M/H] = [M/H]_{\mathrm{GSP-Spec}} - [M/H]_{\mathrm{APOGEE}}$.

We also calibrated the GSP-Spec $\alpha$ abundances on the APOGEE scale.
For stars of spectral type F or cooler, the spectral range covered by the \Gaia Radial Velocity Spectrometer (845 to 872\,nm) is dominated by calcium triplet lines \citep[see e.g. Fig.~17][]{Cropper18}, which means that the \texttt{alphafe\_gspspec} values listed in \Gaia~DR3 are mostly derived from Ca lines. We followed the approach of \citet{RecioBlanco23gspspec}, and calibrated the GSP-Spec $\alpha$ abundances on the APOGEE DR17 [Ca/Fe] abundances rather than [$\alpha$/Fe]. We find that the \texttt{alphafe\_gspspec} values are lower by 0.07\,dex on average (with a dispersion of 0.04\,dex), and that the bias does not correlate with the effective temperature or surface gravity, but depends on metallicity (right panel of Fig.~\ref{fig:calibration_abundances}). We modelled this bias as a linear function of GSP-Spec metallicity,

\begin{equation} \label{eq:correctionALPHA}
    \Delta [\alpha/H] = -4.481 \times 10^{-2} \times  [M/H]_{\mathrm{GSP-Spec}} - 4.414 \times 10^{-2} 
\end{equation}
where $\Delta [\alpha/H] = [\alpha/H]_{\mathrm{GSP-Spec}} - [\alpha/H]_{\mathrm{APOGEE}}$.
We then applied the corrections in equations \ref{eq:correctionMH} and \ref{eq:correctionALPHA} to the entire RC-GSP-Spec sample.

\begin{figure*}
\includegraphics[width=0.99\textwidth]{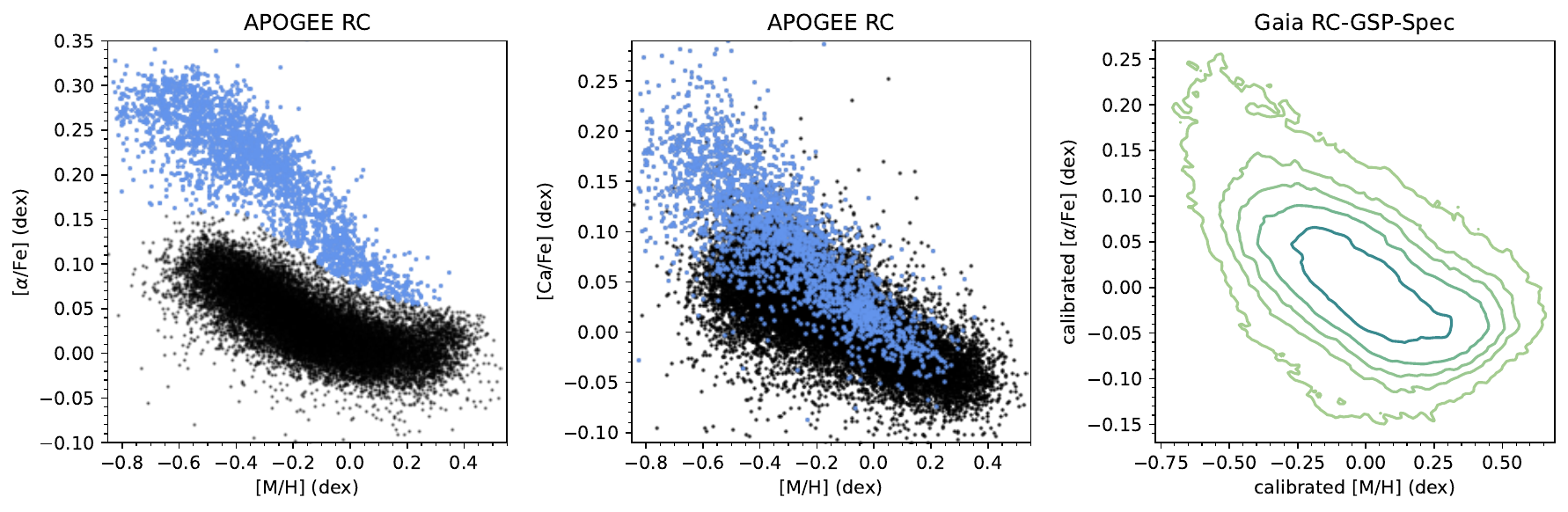} 
\caption{ Red clump stars in chemical abundance space. Left: Distribution of the APOGEE RC stars in the [$\alpha$/Fe]-metallicity plane. The $\alpha$-rich stars are highlighted manually. Middle: Same stars in the [Ca/Fe]-metallicity plane. The [Ca/Fe] abundances do not have the same discriminating power as [$\alpha$/Fe], which accounts for O, Mg, Si, S, and Ca for APOGEE. Right: Distribution of the \Gaia RC-GSP-Spec stars in the [$\alpha$/Fe]-metallicity plane. The contours enclose 99\%, 95\%, 90\%, 80\%, and 50\% of the sample. The \Gaia $\alpha$-abundances are dominated by Ca lines and do not allow us to reliably distinguish the $\alpha$-rich from the $\alpha$-poor disc stars.  \label{fig:alpha_iron_kde}}
\end{figure*}

Figure~\ref{fig:alpha_iron_kde} shows the distribution of APOGEE RC stars in the [$\alpha$/Fe]-metallicity and in the [Ca/Fe]-metallicity planes. The APOGEE [$\alpha$/Fe] (derived as the mean of [O/Fe], [Mg/Fe], [Si/Fe], [S/Fe], and [Ca/Fe]) estimates have a greater discriminating power than [Ca/Fe] alone\footnote{[Mg/Fe] abundances provide the clearest separation between the two sequences \citep{Bensby14, Vincenzo21, Buder21} but cannot be reliably measured by \Gaia.}. The latter barely allows us to distinguish stars between the high- and the low-$\alpha$ disc at low metallicity ([M/H]<-0.3\,dex), and it does this not at all in the metal-rich regime. 

For the rest of this study, we focus on the $\alpha$-poor disc, which we define as the stars with an APOGEE [Ca/Fe] or a calibrated GSP-Spec [$\alpha$/Fe] lower than 0.1\,dex. We discuss the consequences of possible contamination by the thin disc in Sect.~\ref{sec:discussion}.

%
%
\subsection{ Selection function of the \Gaia RC-GSP-Spec sample} \label{sec:gaia_sf}

Building the selection function $S$ for a given sample means identifying the different cuts and selections that cause objects of interest (in the present case, any red clump star in the Milky Way) to be absent from the sample, in order to link the observed number of objects $N_{\textrm{obs}}$ to the true number $N_{\textrm{true}}$, where $N_{\textrm{obs}} = N_{\textrm{true}} \, S$.

If an RC star is not present in the RC-GSP-Spec sample, it can for one of these three reasons:
(i) it is not in the \Gaia DR3 catalogue; (ii) it is in this catalogue, but the selection performed in Sect.~\ref{sec:gaiarc} failed to identify it as an RC star; and (iii) the spectroscopic measurements associated with this star are not of sufficient quality to pass the selections applied in Sect.~\ref{sec:gspspec}. These are independent conditions that result in a product of selection function terms. When we wish to know how many RC stars lie in direction $(\ell,b)$ and at distance $d$, $N_{\textrm{RC,obs}}$, in the sample,
this can be written as

\begin{equation}  \label{eq:N_l_b_d}
    N_{\textrm{RC,obs}}(\ell,b,d) = N_{\textrm{RC,MW}} \times  \frac{ N_{\textrm{RC,Gaia}} }{ N_{\textrm{RC,MW}} } \times  \frac{ N_{\textrm{RC,selected}} }{ N_{\textrm{RC,Gaia}} } \times  \frac{ N_{\textrm{flags}} }{ N_{\textrm{RC,selected}} },
\end{equation}
where $N_{\textrm{RC,MW}}$ is the true number of RC stars at that location in the Milky Way (which we ultimately wish to infer), $N_{\textrm{RC,Gaia}}$ is the number of RC stars at that location that is listed in the \Gaia catalogue, $N_{\textrm{RC,selected}}$ is the number of stars we were able to successfully identify as RC stars, and $N_{\textrm{flags}}$ is the number of RC stars that also satisfy our chosen quality flags. (We omitted the location dependence on the right-hand side of the equal sign in \autoref{eq:N_l_b_d} to simplify the notation.)

We need to express the fraction in terms of direct observables, but the distance $d$ is not an observable. Because RC stars are standard candles with the same intrinsic magnitude $M_G=0.5$\,mag, however, we can rewrite Eq.~\ref{eq:N_l_b_d} as a function of $(\ell, b, G)$. We estimate $G$ at any $(\ell,b,d)$ through the distance $d$ and a 3D extinction map. The completeness is then the product of the above selections,

\begin{equation}   \label{eq:N_l_b_G}
\begin{split}
    \frac{N_{\textrm{RC,obs}}}{N_{\textrm{RC,MW}}}\bigl(\ell,b,G(d,A_G)\bigr)
    &= 
    \underbrace{ S_{Gaia} }_{= 1 \textrm{ if } G<14}
    \,
    \underbrace{ S_{\textrm{RC selection}} }_{= 1 \textrm{ if } G<14}
    \, 
    \underbrace{  \overline{S}_{\textrm{RC-GSP-Spec}} }_{= 0 \textrm{ if } G>14}
\end{split},
\end{equation}
where we have used $S$ as a shorthand for the ratios in Eq.~\ref{eq:N_l_b_d}, and all the terms on the right-hand side are functions of $(\ell,b,G)$; and $G(d,A_G)$ is shorthand for $G\bigl(d,A_G(\ell,b,d)\bigr)$.  We elaborate on the symbol choice of $\overline{S}_{\textrm{RC-GSP-Spec}}$ in Sect.~\ref{sec:SF_met}.

In Eq.~\ref{eq:N_l_b_G}, the term $S_{Gaia}$ is 1 if $G$$<$14 because the overall \Gaia catalogue is essentially complete at these bright magnitudes \citep[][]{CantatGaudin23}. $S_{\textrm{RC selection}}$ is also 1 in this magnitude range because the parallax errors are sufficiently small to enable the selection of virtually all red clump stars (see Sect.~\ref{sec:gaiarc} and Fig.~\ref{fig:gaia_rc_selection}). The last term in this equation describes the fraction of identified RC stars that also satisfy the quality flags we imposed on the RC-GSP-Spec sample. We remark that none of our final RC-GSP-Spec sample stars is fainter than $G$=14\,mag. Since the completeness of the sample is 0\% for $G$>14\,mag (setting the selection function to zero), we only need to describe the behaviour of the other terms in the brighter regime. We construct $S_{\mathrm{RC-GSP-Spec}} (\ell,b,G)$ in Sect.~\ref{sec:compute_S_ratios}.

\subsubsection{The \textit{Gaia} selection function dependence on magnitude and sky position} \label{sec:compute_S_ratios}

We investigated the completeness $S_{\mathrm{RC-GSP-Spec}} (\ell,b,G)$ of the RC-GSP-Spec sample (constructed in Sect.~\ref{sec:gspspec}) with respect to all 1.8M \Gaia DR3 red clump stars, which we assumed to be fully complete in the magnitude range in which precise GSP-Spec abundances are available ($G$<14). Computing the selection function as ratios with respect to a more complete sample is a common approach in astronomy \citep[see e.g.,][for applications to \Gaia data]{Rybizki21rvs, CastroGinard23_subsamples}.

\begin{figure}
\includegraphics[width=0.99\columnwidth]{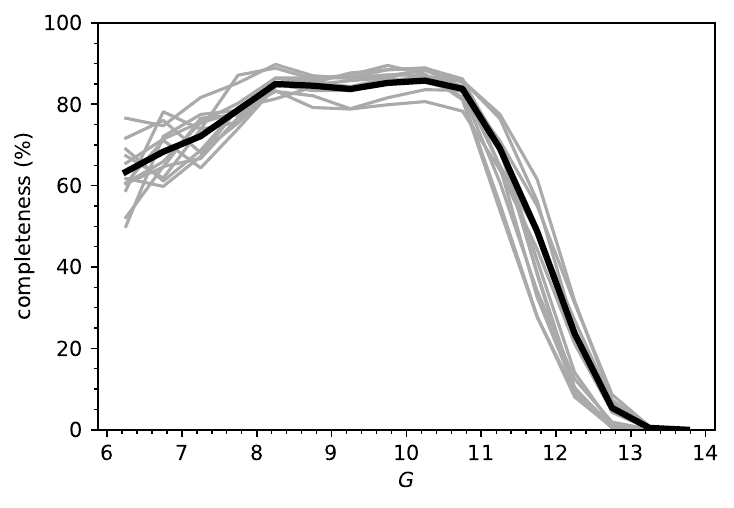} 
\caption{ Completeness of our RC-GSP-Spec sample with respect to all \Gaia DR3 red clump stars computed in bins of 0.5\,mag (black line). The grey lines, computed independently in the 12  HEALPixels at level 0, show the same quantity. \label{fig:completenes_rc-gsp-spec}}
\end{figure}

In order to understand the general behaviour of the selection function with $G$, we first computed the count ratios in bins of 0.5\,mag computed over the entire sky and then in HEALPix regions at level 0, which correspond to 12 areas of about 3400\,deg$^2$ each\footnote{The HEALPix scheme \citep{2005ApJ...622..759G} allows to tessellate the celestial sphere in $12 \times 4^N$ equal-area regions, where $N$ is the level of subdivision.}. We show both cases in Fig.~\ref{fig:completenes_rc-gsp-spec}.
The completeness is close to 80\% for sources with $8 < G < 10.5$\,mag, and it then decreases to zero near 13.5 magnitudes. The completeness also decreases slightly for $G<8$\,mag. Because the parallax correction of \citet{Lindegren21parallaxbias} we employed in Sect.~\ref{sec:gaiarc} is not available for stars brighter than $G=6$\,mag, our completeness is zero at the bright end (but only 0.05\% of the sources with available GSP-Spec abundances are brighter than $G$=6).

We aimed to construct a selection function that reflects both the spatial and the magnitude dependence of the completeness.
We explored different binning schemes to find a compromise between spatial resolution, magnitude resolution, and reasonable noise levels. Because the magnitude range 6<$G$<8\,mag only contains 2\% of the RC-GSP-Spec sample, we grouped sources with 6<$G$<10.5\,mag in a single magnitude bin and split the rest of the sample into bins of 0.5\,mag. We tessellated the sky into HEALPix regions at level 3 (about $7.4$\,deg$^2$). Although the fine details of the \Gaia scanning pattern are better visualised with a finer spatial resolution, increasing the HEALPix level would produce noisier maps or require the use of broader magnitude bins (degrading the distance resolution of the selection function).

\begin{figure*}
\includegraphics[width=0.99\textwidth]{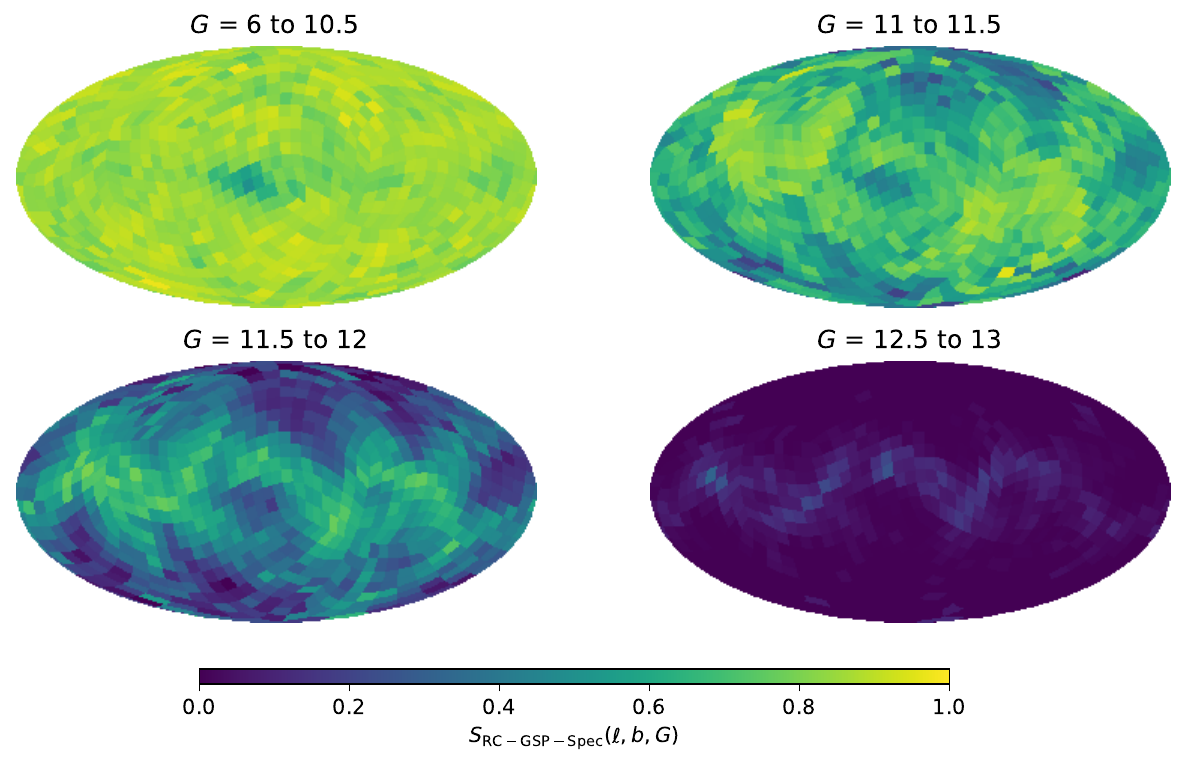} 
\caption{  Selection function of the RC-GSP-Spec sample, constructed from count ratios in spatial (healpix level 3) and magnitude bins. The maps are in Galactic coordinates, centred on $(\ell,b)$=(0,0), with $\ell$ increasing to the left. \label{fig:gaia_chem_SF_allsky_fourpanels}
}
\end{figure*}

Fig.~\ref{fig:gaia_chem_SF_allsky_fourpanels} shows all-sky completeness maps in four magnitude intervals. Even at the bright end, \Gaia is not complete. It is especially visible around $(\ell,b)$=(5,-5) in the top left panel of Fig.~\ref{fig:gaia_chem_SF_allsky_fourpanels}, which contains a darker area of a few degrees below and to the east of the Galactic centre (around $(\ell,b)$=(5,-5)). This region (which includes Baade's Window) corresponds to a direction with particularly low extinction, which leads to an effectively high \Gaia source density near the crowding limit of the survey. The source density is so high that it affects the completeness of the \Gaia RVS catalogue itself \citep[see Figure~2 in][]{CantatGaudin23}. 
Patterns appear in our map for the magnitude range $11 - 11.5$\,mag. These correspond to the imprint of the \Gaia scanning law and appear as stripes of lower completeness, mostly caused by the \texttt{ruwe}<1.4 filter applied to the \citet{GaiaDR3chemcartoPVP} GSP-Spec sample. We discuss this bias in Appendix~\ref{app:ruwe}. At $G$>11.5\,mag, the more complete regions correspond to areas that the \Gaia satellite observed more frequently (more visits; bottom panels of Fig.~\ref{fig:gaia_chem_SF_allsky_fourpanels}).

Another possible approach that we did not follow here would have been to construct an analytic function of basic observables \citep[][for instance, only use the $G$ magnitude]{Rix21} to predict the probability that a given source passes the selection filters. This approach is simple and powerful when the effect of cuts on,measurement errors or parallax signal-to-noise ratio is to be modelled, for example, which are mostly smooth functions of $G$ in any given direction. The approach would also allow us to construct a selection function whose arguments are observable quantities even when they are not always available for all sources. The drawback of this approach is that it requires the user to accurately describe every single cut to the data with an appropriate model. In the present case, this would mean describing the mentioned \texttt{ruwe} quirk near $G\sim11.5$ (described in Appendix~\ref{app:ruwe}) as a function of position and magnitude, as well as the various quality cuts applied to the GSP-Spec sample by \citet{GaiaDR3chemcartoPVP}. Our chosen approach allows us to compute $S_{\textrm{RC-GSP-Spec}}$ in a single ratio step.

\subsubsection{Effective red clump star selection function in Galactic coordinates } \label{sec:gaiachem_eff_sf}

The selection function $S_{\mathrm{RC-GSP-Spec}}$ we constructed in Sect.~\ref{sec:gaia_sf} is a function of sky coordinates and apparent $G$ magnitude. In the context of mapping the Milky Way disc, we are generally interested in knowing the completeness of a given sample as a function of spatial Galactic coordinates.

The apparent magnitude $G$ of a star at any location in the Milky Way can be expressed as

\begin{equation}
    G(\ell,b,d) = M_G + 5 \log_{10} d - 5 + A_G (\ell,b,d,T_{\mathrm{eff}}) \label{eq:G},
\end{equation}

\noindent where $M_G$ is the absolute magnitude of the star, $d$ is its distance (in parsecs), and $A_G$ is the total amount of interstellar extinction in the $G$ band to the star, which we could estimate from a 3D extinction map and the effective temperature $T_{\mathrm{eff}}$ of that star\footnote{Formally, obtaining $A_G$ from the $A_0$ provided by extinction maps depends on the entire energy distribution over the $G$ band, but for red clump stars the effects of metallicity and surface gravity are negligible compared to effective temperature.}. 
For any particular choice of a star (defined by a value of reddening-free magnitude $G_0$ and temperature $T_{\mathrm{eff}}$), we can use equation~\ref{eq:G} to turn a function of $(\ell,b,G)$ into the effective selection function $S_{\mathrm{RC-GSP-Spec}} (\ell,b,d)$.
We refer to Section~3 of \citet{Bovy17} for an illustration of how a given selection function translates into a different effective selection function for different stellar populations in the Tycho-\Gaia Astrometric Solution catalogue \citep[][]{Michalik15}. 

\begin{figure*}
\includegraphics[width=0.99\textwidth]{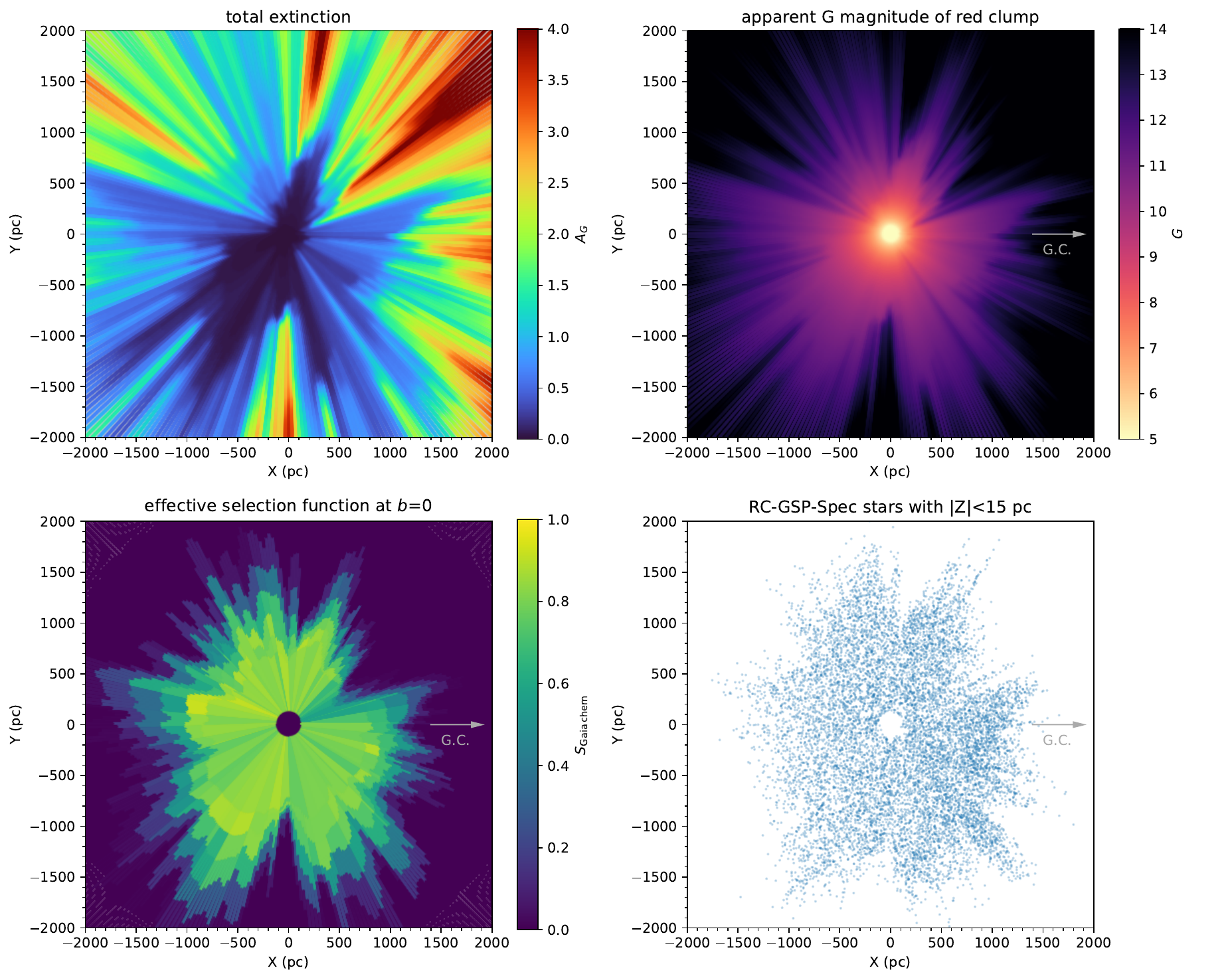} 
\caption{ Effect of distance and extinction on the sample completeness. Top left: Cumulative extinction at $Z=0$, computed from the \citet{Lallement22} dust map. Top right: Predicted apparent $G$ magnitude for a red clump star, assuming an intrinsic $M_G$=0.5\,mag. Bottom left: Selection function $S_{\mathrm{Gaia\,chem}}(X,Y,Z)$ plotted for $b$=0. Bottom right: $XY$ distribution of red clump stars with $|Z|$<15\,pc in our RC-GSP-Spec sample (Sect.~\ref{sec:gspspec}). The Galactic centre is to the right in every panel.\label{fig:gaia_chem_SF_XY}
}
\end{figure*}

In this paper, we study the spatial distribution of red clump stars, for which we assumed $M_G$=0.5\,mag \citep[][]{RuizDern18,Hall19} and $T_{\mathrm{eff}}$=4700\,K \citep[][]{Girardi16}. We used the all-sky dust maps of \citet{Lallement22}, and converted their extinction $A_0$ into $A_G$ using the precomputed values provided by the Python package \texttt{dustapprox}\footnote{\texttt{dustapprox}: \url{https://github.com/mfouesneau/dustapprox}} for this effective temperature. Fig.~\ref{fig:gaia_chem_SF_XY} shows the expected $G$ magnitude and the corresponding effective selection function for red clump stars in the Galactic mid-plane.

We find no overall systematics between the observed $G$ magnitudes and those predicted by the dust extinction model, except in a few distant regions (>2\,kpc from us) where the \citet{Lallement22} map allows for zero extinction, and the observed red clump stars can be fainter than predicted by $\sim$0.1\,mag. In high-extinction regions, some stars are slightly brighter than predicted by the model. This effect may be caused by the spatial resolution of the map, which smooths out the fine structure of the dust clouds at scales smaller than 5\,pc. Because the selection function drops from 80\% to 0\% over the course of two magnitudes (Fig.~\ref{fig:completenes_rc-gsp-spec}) and is constructed in ranges of 0.5 in magnitudes, we did not consider these 0.1\,mag biases (which only affect sparsely populated regions) to have a significant impact.

\subsubsection{Metallicity dependence of the selection function} \label{sec:SF_met}

\begin{figure*}
\includegraphics[width=0.99\textwidth]{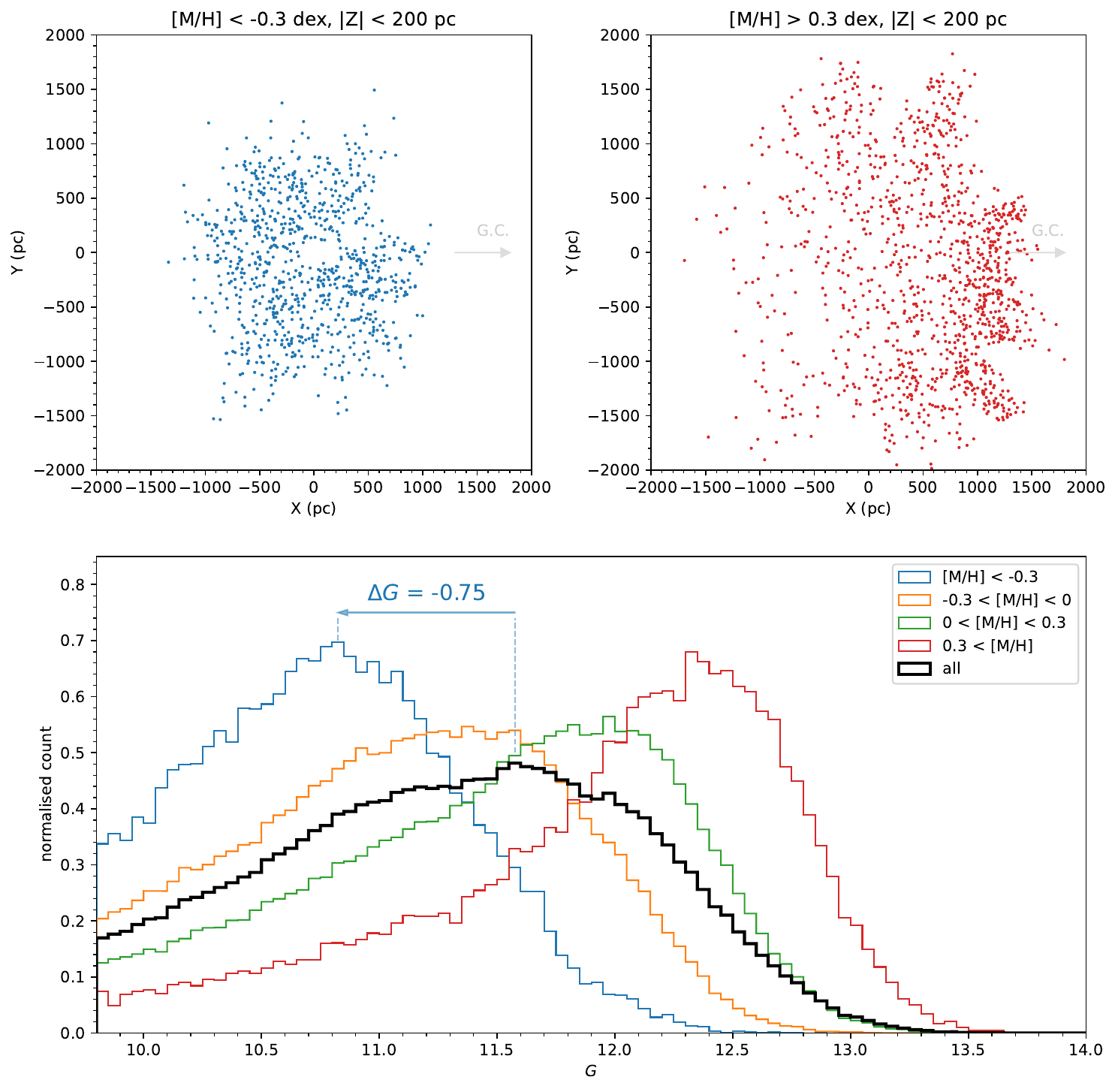} 
\caption{  Effect of metallicity on the selection function. Top left: Distribution of metal-poor RC-GSP-Spec stars projected the Galactic $(X,Y)$ plane for stars with $|Z|$<200\,pc alone. The Sun is at $(X, Y)$=(0,0), and the Galactic centre is to the right. Top right: Same for the most metal-rich stars in the sample. Metal-rich stars contained in the sample reach larger distances from the Sun. Bottom: $G$ magnitude distribution in the entire sample (black) and in four ranges of metallicity. The offset $\Delta G$ used in equation~\ref{eq:N_l_b_d_FeH} is indicated for the most metal-poor metallicity group. \label{fig:sf_with_metallicity}}
\end{figure*}

One of our quality flags for inclusion in the RC-GSP-Spec sample is that the combined metallicity and alpha-abundance uncertainty of a star be smaller than 0.1\,dex (Eq.\,\ref{total_uncert_req}). This depends foremost on the stellar brightness, but at a given brightness also on the stellar metallicity.  Hence, our selection not only favours brighter sources and regions of the sky that were more frequently visited by \Gaia (Fig.~\ref{fig:gaia_chem_SF_allsky_fourpanels}), but also favours high-metallicity stars whose spectral lines are inherently more prominent and yield smaller uncertainties. As a result, metal-poor stars included in the RC-GSP-Spec sample tend to be slightly brighter and more nearby than metal-rich stars, as Fig.~\ref{fig:sf_with_metallicity} illustrates.

The selection function $\overline{S}_{\textrm{RC-GSP-Spec}}$ shown in Figs.~\ref{fig:gaia_chem_SF_allsky_fourpanels} and \ref{fig:gaia_chem_SF_XY} only represents the metallicity-averaged selection function for the whole RC-GSP-Spec sample (black line in the lower panel of Fig.~\ref{fig:sf_with_metallicity}); hence the choice of $\overline{S}$.  In practice, the apparent $G$-magnitude distribution of stars in the RC-GSP-Spec sample depends quite strongly on metallicity, as the coloured lines for different metallicity subsets in Fig.~\ref{fig:sf_with_metallicity} show. These differences are most likely mostly due to differences in  $S_{\mathrm{RC-GSP-Spec}} (G)$, not any intrinsic differences in the spatial distribution. 
We tookan empirical approach to account for this metallicity dependence of the selection function.  We did this by assuming that for any metallicity, the selection function is well described by the metallicity-averaged function (black line in Fig.~\ref{fig:sf_with_metallicity}), just shifted by the amount by which the peaks in the metallicity sub-bins differ from the peak in the metallicity-averaged distribution. For instance, for the metal-poor sub-sample ([M/H]$<-0.3$), we shifted the mean selection function so that it was 0.75 magnitudes brighter.

Expressed mathematically, we did this via
\begin{equation}\label{eq:N_l_b_d_FeH}
    S_\textrm{RC-GSP-Spec}(\ell,b,G,[M/H])\equiv \\ \overline{S}_{\textrm{RC-GSP-Spec}}\bigl (\ell,b,G-\Delta G\bigr ),
\end{equation}
where $\Delta G = \Delta G([M/H])$ is the metallicity-dependent shift of the peaks in the distributions in the bottom panel of Fig.~\ref{fig:sf_with_metallicity}.
We resorted to this empirical approach because metallicities are not available for all stars in the \Gaia catalogue, which would be needed to express the selection function as a ratio in the style of Eq.~\ref{eq:N_l_b_d}.

This approach could be refined even further by accounting for the fact that RC stars are not perfect standard candles and that their absolute magnitudes vary with age and metallicity \citep[from $M_G$$\sim$0.35 for our most metal-poor stars to $\sim$0.6 according to PARSEC isochrones;][]{Bressan2012MNRAS.427..127B}. The mean absolute magnitude also depends on the age distribution within each metallicity group, with older RC stars being slightly fainter. Second-order corrections beyond the approach used in this study would therefore require deeper assumptions on the age-metallicity distribution at any point in the Milky Way. The overall effect of this finer correction would be to reduce the estimated effective volume of the metal-poor stars by $\sim$0.1 in distance modulus (because they are brighter, a given observed $G$ corresponds to a shorter distance), and to increase the effective volume of the metal-rich group by a similar amount\footnote{In other words: if all $G$ distributions in the bottom panel of Fig.~\ref{fig:sf_with_metallicity} overlapped perfectly, it would mean that the metal-poor sample is deeper than the metal-rich sample.}, which we expect would have a negligible impact on the overall results of this study. We point out that this effect is present in the construction of the APOGEE selection function as well because the absolute $H$-band magnitude of RC stars also depends on their age and metallicity.

%
%
\section{Modelling the Galactic disc profile for mono-abundance populations} \label{sec:density}

In this section, we first qualitatively visualise the density profile of the disc (Sect.~\ref{sec:qualitative}). This simple approach is statistically unsound, however. In Sect.~\ref{sec:density_model} we build a forward model of the spatial distribution that accounts for the selection functions of both \textit{Gaia} and APOGEE, and we provide quantitative results.

\subsection{Qualitative \Gaia density profile in the solar neighbourhood} \label{sec:qualitative}

We focused on the $\alpha$-poor stars ([$\alpha$/Fe]<0.1\,dex) in the RC-GSP-Spec sample. 
We binned them into six bins of [M/H], and a 200\,pc grid in $(X,Y,Z)$ coordinates and computed the mean selection function in each bin (by sampling it on a much finer grid and then averaging) to estimate the completeness. We obtained corrected counts in each bin by dividing the observed counts by the completeness.

To visualise the radial density profile in each metallicity bin, we plot the resulting density as a function of Galactocentric distance $R$ for the bins closest to the Galactic mid-plane (-200<$Z$<200\,pc), assuming $R_\odot=8.15$\,kpc \citep[][]{Reid19}.

\begin{figure}
\includegraphics[width=0.99\columnwidth]{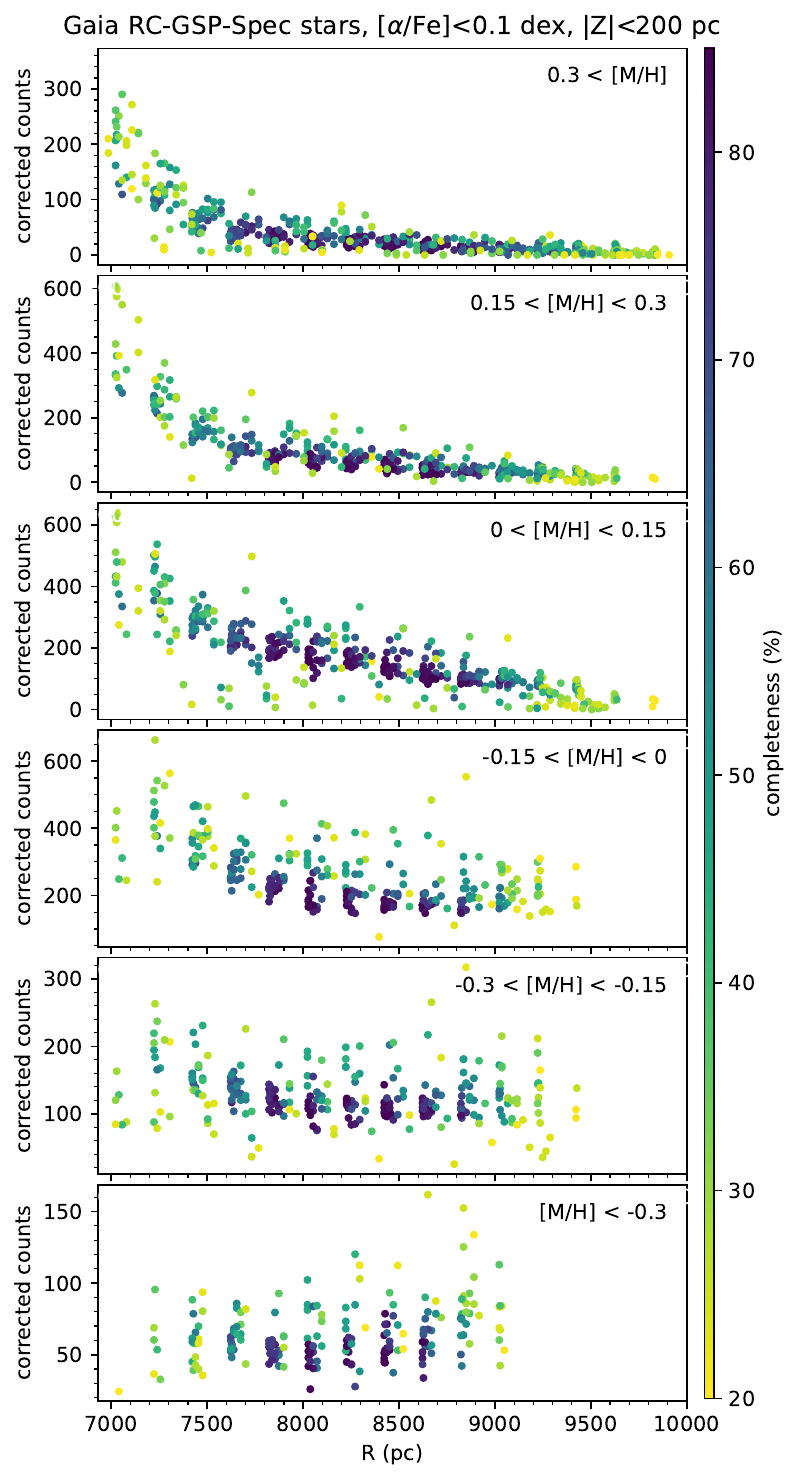} 
\caption{Completeness-corrected number of $\alpha$-poor RC-GSP-Spec stars in bins of Galactic (X, Y) bins of 200$\times$200$\times$200\,pc near the Galactic mid-plane (-200<$Z$<200\,pc) as a function of Galactocentric distance in parsecs. We selected $\alpha$-poor stars as [$\alpha$/Fe]<0.1\,dex. A dot represents the corrected density for regions in which the expected completeness is greater than 20\%. The top and bottom panels correspond to the data displayed in the top left and top right panels of Fig.~\ref{fig:sf_with_metallicity}, respectively. 
We stress that the observed counts should only be divided by completeness for qualitative visualisation and data exploration purposes. Section~\ref{sec:modelfitting} implements a forward-modelling approach to apply the selection function to a density model. \label{fig:gaia_density_corrected}
}
\end{figure}

Figure~\ref{fig:gaia_density_corrected} shows the resulting completeness-corrected Galactocentric distance distributions per [M/H] bin for the stars closest to the Galactic mid-plane (-200<$Z$<200\,pc).
The density increases towards the Galactic centre in all [M/H] bins, except for the most metal-poor ([M/H]<-0.3\,dex) population, and the slope steepens with metallicity. The corrected density is noisier in regions of low completeness (darker points in Fig.~\ref{fig:gaia_density_corrected}) because they include fewer stars and because of the uncertainties on the selection function itself.

An example of the vertical density profile traced by corrected counts is shown in Appendix~\ref{app:hZmetalrich} for the metal-rich ([M/H]>0.3\,dex) population. In the rest of this section, we fit a density model to the full 3D distribution of stars.

\subsection{\Gaia and APOGEE: Tracing the large-scale structure of the disc} \label{sec:density_model}

In this section we use a forward-modelling approach to fit an analytical density model to the observed stellar distribution.

\subsubsection*{Density model}

We describe the stellar density distribution $\nu_\star$ in the Milky Way disc with the model used by \citet{Bovy16population} and \citet{Mackereth17}, which assumes that the density is axisymmetric and that the radial and vertical density profiles are separable,
\begin{equation}
\nu_\star(R, \phi, Z) = \nu_\star(R, Z) =  \Sigma(R) \,  \zeta(Z|R) \quad, \text{where } \int \zeta(Z|R) dZ = 1 .
\end{equation}

In this model, we assume that the radial density profile is a radially broken exponential,
\begin{equation}
\ln \Sigma(R) \propto  \begin{cases}
    -h_{R,\text{in}}^{-1}(R-R_\odot)  & \quad, \text{where } R \leq R_{\text{peak}}\\
   -h_{R,\text{out}}^{-1}(R-R_\odot)  & \quad, \text{where } R > R_{\text{peak}}\\
  \end{cases},
\end{equation}
where $R_\odot=8.15$\,kpc is the solar Galactocentric radius \citep{Reid19}, $R_{\text{peak}}$ is the radius at which the change of slope occurs, and $h_{R,\text{in}}$ and $h_{R,\text{out}}$ are the scale length inside and outside the break radius, respectively. The density continuity at $R=R_\odot$ sets the relative normalisation of the two parts.

We assumed that the vertical profile is single exponential with a flare, making the scale height itself an exponential function of the Galactocentric distance $R$,
\begin{equation} \label{eq:hZ}
h_Z^{-1}(R) = h_{Z,\odot}^{-1} \exp\left[R_{\mathrm{flare}}^{-1}(R-R_\odot)\right],
\end{equation}
with $h_{Z,\odot}$ the scale height at the location of the Sun, and $R_{\mathrm{flare}}$ the flaring scale length. A negative value of $R_{\mathrm{flare}}$ corresponds to a $h_Z^{-1}$ decreasing with $R$, that is, a scale height $h_Z$ increasing at larger radii. We illustrate the impact of flaring on the vertical structure in Appendix~\ref{app:hZmetalrich}.
The (normalised) vertical density profile $\zeta$ at any $R$ is given by

\begin{equation}
\ln \zeta(Z|R) = - h_Z^{-1} ~ |Z| - \ln{2h_Z(R)}.
\end{equation}

\subsubsection*{APOGEE and \Gaia joint selection function}

The APOGEE targets are chosen by randomly picking sources from the 2MASS catalogue, while the RC-GSP-Spec (\Gaia) data do not rely on another input catalogue for its target selection or data processing. We can therefore consider the two samples as independent. According to set theory, the probability that a given RC star ends up in either or both of the catalogues is thus obtained as the sum, minus the probability that it is present in both,

\begin{equation} \label{eq:Scombined}
\begin{split}
S_{\mathrm{combined}} = & S_{\mathrm{RC-GSP-Spec}} 
                           +  S_{\mathrm{APOGEE}} \\
                        & - S_{\mathrm{RC-GSP-Spec}} \times  S_{\mathrm{APOGEE}}, 
\end{split}
\end{equation}

\noindent where all terms are functions of $(\ell,b,d)$ (or, equivalently, $X$, $Y$, and $Z$), expressing the selection probability at any given location in the Milky Way, and $S_{\mathrm{RC-GSP-Spec}}$ also depends on metallicity (Eq.~\ref{eq:N_l_b_d_FeH}).

We stress that if the APOGEE survey assigned higher observing priorities to targets with available \Gaia data, or if \Gaia relied on APOGEE data to refine its GSP-Spec values, the assumption of independence would not hold, and it would be more complicated to calculate the joint selection function. For instance, \citet{Rybizki21rvs} have shown that in \Gaia~DR2, the selection function of the Radial Velocity Sample was strongly shaped by the Initial \Gaia Source List \citep[IGSL;][]{Smart14} and is therefore correlated with the selection function of SDSS DR9 \citep{York00,Ahn12} and GSC2.3 \citep{Lasker08} from which the IGSL was built. This issue was solved in the \Gaia~DR3 catalogue because the processing of the Radial Velocity Sample no longer relies on literature catalogues.

As \Gaia has quickly become the reference catalogue for identifying objects of interest (from Milky Way stars to distant quasars), the next generation of large-scale spectroscopic surveys such as WEAVE \citep{Dalton14, Jin23} or 4MOST \citep{deJong12, deJong14} will produce catalogues whose selection function will correlate strongly with that of \Gaia, and in certain cases, they will even simply constitute subsamples of the \Gaia data.

\begin{figure}
\includegraphics[width=0.99\columnwidth]{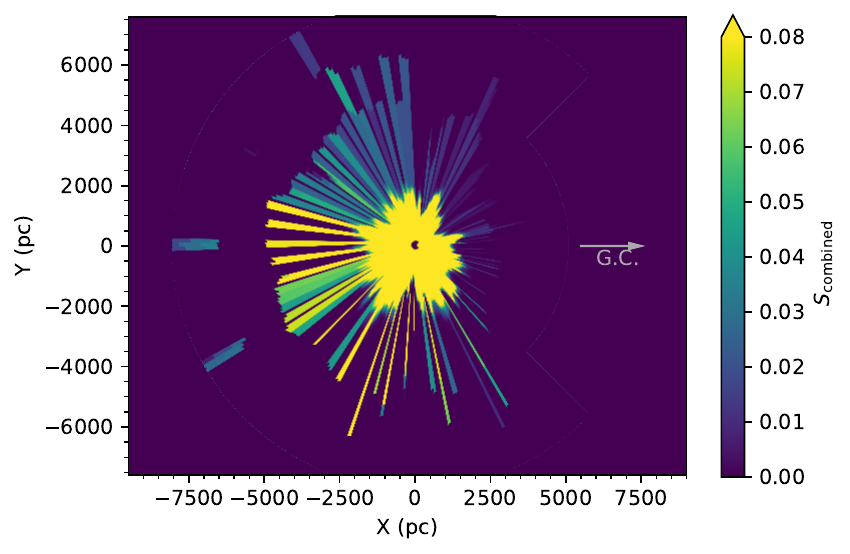} 
\caption{Effective selection function (shown at $b$=0$^{\circ}$) for the combination of our \Gaia and APOGEE samples obtained from the effective selection functions displayed in Figs.~\ref{fig:apogee_sf_xy} and \ref{fig:gaia_chem_SF_XY} and equation~\ref{eq:Scombined}. The completeness reaches over 80\% in the solar neighbourhood dominated by the \Gaia sample. \label{fig:combined_sf}}
\end{figure}

\subsection{Model fitting} \label{sec:modelfitting}

For a given choice of model parameters 
$$\theta = (h_{R,\text{in}},\, h_{R,\text{out}},\, R_{\text{peak}},\, h_{Z,\odot} ,\, R_{\mathrm{flare}}^{-1}  ), $$ the model density at location $(X,Y,Z)$ is $\nu_\star (X,Y,Z|\theta)$, and thus, the predicted observed number of stars is

\begin{equation}
N_{\mathrm{pred}}(X,Y,Z|\theta) = N_0 \, \nu_\star (X,Y,Z | \theta ) \, S(X,Y,Z), 
\end{equation}
where $S$ represents the selection function, which is the APOGEE selection function (see Sect.~\ref{sec:apogee_sf}), the RC-GSP-Spec metallicity-dependent selection function (see Sect.~\ref{sec:SF_met}), or the combined one (see Eq.~\ref{eq:Scombined}). We fixed $N_0$ for each model so that the total number of predicted counts matched the total number of observed stars in the bins where $S$>0.

We fit the parameter set $\theta$ by maximising the Poisson likelihoods of the observed stellar counts with those predicted by our model per spatial bins $(X, Y, Z)$,  
\begin{equation}
\log \mathcal{L}(\theta) =  \sum_{X, Y, Z} \log P \left(  N_{\mathrm{obs}}(X, Y, Z) , N_{\mathrm{pred}}(X, Y, Z\,|\,\theta) \right),
\end{equation}
where the $(X, Y, Z)$ bins are $200 \times 200 \times 200$\,pc cubes. To avoid biases due to a possibly inaccurate extrapolation of the \citet{Lallement22} extinction map towards the inner disc, we also limited the fitting to bins with $R$>4\,kpc (this only discarded 2\% of the APOGEE stars and none of the \Gaia stars).

We sampled the log-posterior with the Markov chain Monte Carlo sampler \texttt{emcee} \citep{ForemanMackey13}, exploring the parameter space for 1500 steps with 16 walkers and imposing very broad flat priors.

\begin{figure*}
\includegraphics[width=0.99\textwidth]{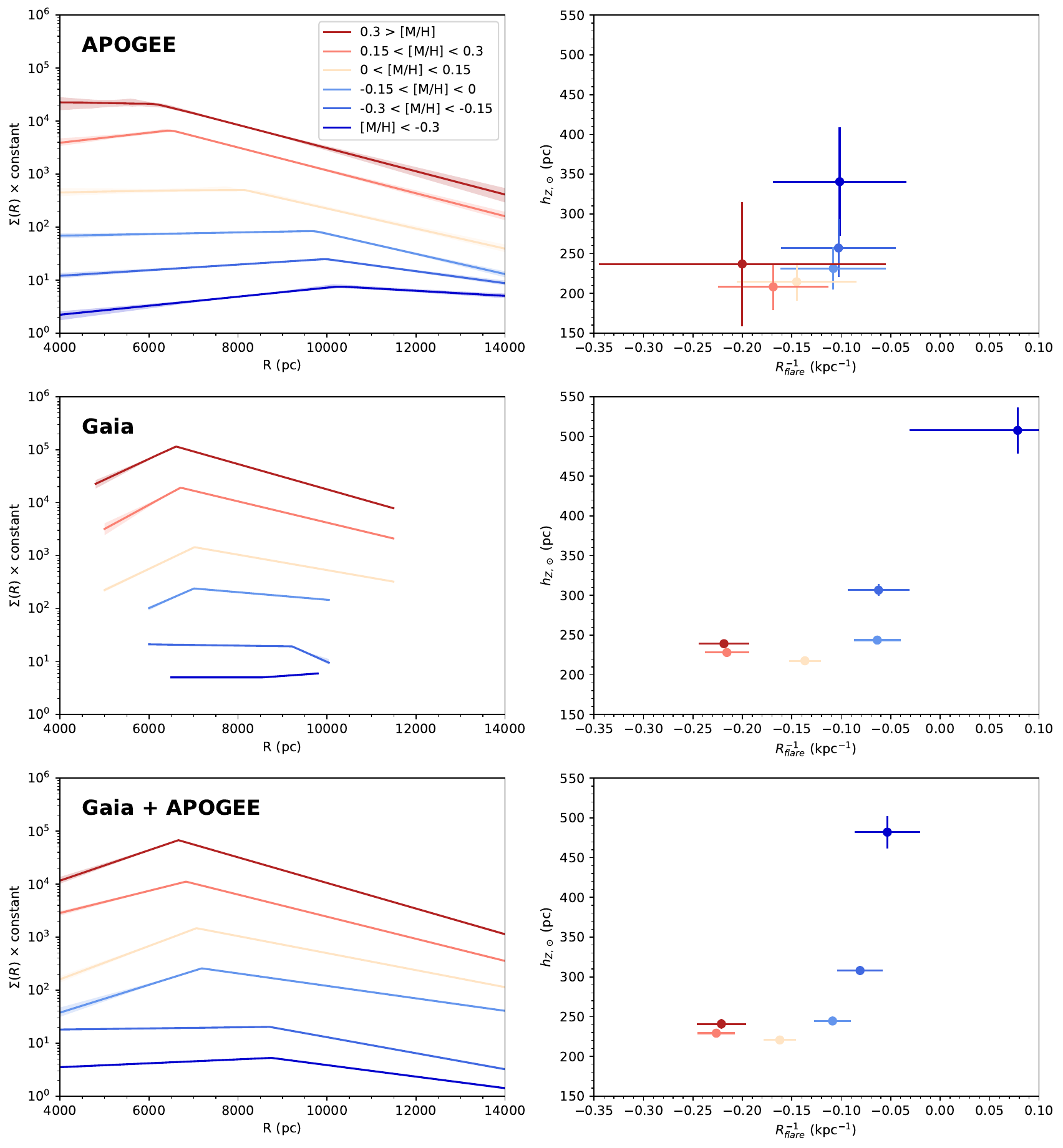} 
\caption{Density profiles of red clump stars from APOGEE, \Gaia RC-GSP-Spec, and their combined data sets (from top to bottom).  The left panels show the radial density profiles $\Sigma(R)$ obtained in six ranges of metallicity in the $\alpha$-poor disc ([$\alpha/Fe]<0.1$\,dex). The profiles are shifted by an arbitrary offset to be visually separated. The panels on the right represent the flaring scale length and vertical scale height predicted by our model for each metallicity bin.
The shaded areas (left) and error bars (right) represent the 95\% confidence interval. The \Gaia RC-GSP-Spec star sample covers a smaller range in $R$, especially at low metallicity. We list the best-fit parameters in Table~\ref{tab:mcmc}.
\label{fig:density_profiles_all}
}
\end{figure*}

We performed the fitting procedure three times: (i) Using only the APOGEE RC stars, (ii) using only the \Gaia RC-GSP-Spec stars, and (iii) using the combined sample. Fig.~\ref{fig:density_profiles_all} shows the resulting density profiles. In all cases, we discarded stars whose APOGEE [Ca/Fe] or \Gaia GSP-Spec [$\alpha$/Fe] estimates were greater than 0.1\,dex to focus on the $\alpha$-poor disc. 
We discuss these results in Sect.~\ref{sec:discussion} and compare them to previous studies of the density profile of mono-abundance populations in the Milky Way.

We also performed the analysis for the APOGEE stars by selecting the $\alpha$-poor stars based on their APOGEE [$\alpha$/Fe]. The results were very similar, suggesting that we are not strongly biased by our inability to separate the $\alpha$-rich disc (whose relative weight is lower) from the $\alpha$-poor disc.

To assess the impact of the uncertainty on the selection function, we performed the same fitting procedure on the \Gaia RC-GSP-Spec sample restricted to regions in which the sample is at least 80\% complete. The results are shown in Appendix~\ref{app:80pct} (Fig.~\ref{fig:density_gaia_80}). This restriction reduces the radial coverage of the data greatly (to the solar neighbourhood), whicgh increases the parameter uncertainties. However, the radial density profiles and vertical scale lengths are statistically identical to those obtained with the entire RC-GSP-Spec sample.

%
%
\section{Discussion} \label{sec:discussion}

The results we obtained from APOGEE DR17 data (top row of Fig.~\ref{fig:density_profiles_all}) are overall consistent with those of \citet{Bovy16population} and \citet{Mackereth17}, who both used APOGEE DR12 data. In this section, we discuss the difference between the radial density profiles obtained with APOGEE and \Gaia, which we attribute to their very different sampling of the disc. We also discuss the improved description of the vertical density profile obtained by combining the two surveys. Finally, we discuss some remaining sources of uncertainty that are not currently accounted for in our error budget.

\subsection{Radial density profile}

The radial profiles traced by APOGEE data peak at a Galactocentric distance $R_{\text{peak}}$ that continuously increases with decreasing metallicity. 
The radial density distribution traced by the \Gaia RC-GSP-Spec sample (middle row of Fig.~\ref{fig:density_profiles_all}) appears to be broadly consistent, but with some major differences. The radial density profiles of the two populations with [M/H]<-0.15 appear mostly flat, which we attribute to the much shorter coverage of this sample, especially for low metallicities. 
All other populations peak at $R_{\text{peak}}\sim7$\,kpc, which is at odds with the APOGEE results, where the near-solar metallicity samples peak between 8 and 10\,kpc. The density profiles shown in Fig.~\ref{fig:gaia_density_corrected} indicate that in this metallicity range, the density clearly increases with decreasing $R$. We attribute the difference between the profiles obtained from the APOGEE and \Gaia data to their very different spatial coverage. Figure~\ref{fig:blue_orange_panels} shows that APOGEE mostly covers the disc outside the solar neighbourhood, while the \Gaia RC-GSP-Spec sample provides a dense coverage for 7<$R$<9\,kpc. 

It is also likely that the true density profile of the $\alpha$-poor population is not perfectly smooth (owing, e.g., to perturbations by the spiral arms and other non-axisymmetric features). The fit to the APOGEE data therefore describes the overall density distribution over a large scale, with a focus on the disc outside the solar radius, while \Gaia provides stronger constraints on the local density.
\citet{Mackereth17} reported that their best-fit model to the APOGEE data underpredicts the stellar density in the solar neighbourhood compared to \citet{McKee15}, for example. The difference between APOGEE predictions for the solar neighbourhood and other data sets could be resolved with a more flexible (but more complex) model, which would allow us to capture the detailed small-scale variations revealed by \Gaia in addition to the general trends traced by APOGEE.

\subsection{Vertical density profile}

The shallower depth of the \Gaia RC-GSP-Spec sample is counterbalanced by its nearly isotropic coverage, while the deeper APOGEE footprint is mostly dominated by low-latitude fields. The \Gaia RC-GSP-Spec provides better leverage than APOGEE for a constraint of the local vertical scale height of the disc ($h_{Z,\odot}$). The uncertainties on the scale heights obtained from the \Gaia sample are smaller by two to five times than those obtained with APOGEE, and they confirm previous reports that the scale height in the solar neighbourhood decreases with increasing metallicity \citep[e.g.][]{Bovy16population, Yu2021ApJ...912..106Y, Lian2022MNRAS.513.4130L}.
The only population for which \Gaia and APOGEE return discrepant values is the most metal-poor group, for which APOGEE suggests $h_{Z,\odot}$=340$\pm$70\,pc, while \Gaia indicates 507$\pm$30\,pc. This difference is likely due to contamination by the $\alpha$-rich disc, whose scale height is significantly larger \citep[$\sim$1\,kpc according to][]{Bovy16population}, but whose presence in the outer disc (and therefore in the APOGEE sample) is less significant due to its shorter scale length.

An unexpected result driven by the \Gaia data is that the flaring of the $\alpha$-poor (young) disc appears to increase with metallicity, with $R_{\mathrm{flare}}^{-1}$ ranging from $\sim$-0.05 to -0.20\,kpc$^{-1}$ within the metallicity range of our sample. In other words, the scale height $h_Z$ of the more metal-rich stellar populations increases more strongly with $R$ (Eq.~\ref{eq:hZ}). This effect has never been reported in the literature to our knowledge.
The large uncertainties on $R_{\mathrm{flare}}^{-1}$ obtained from APOGEE \citep[in this study and in][]{Bovy16population} do not allow us to draw conclusions on the relation between metallicity and flaring. However, splitting the disc into age groups, \citet{Mackereth17} and \citet{Anders2023A&A...678A.158A} reported that the youngest populations, which exhibit the smallest scale heights, are also more strongly flared. This is expected if the flaring is due to external perturbations because in this scenario, the scale height of the perturbed populations become asymptotically similar in the outer disc, so that populations that are geometrically thinner in the inner disc are more strongly flared by definition. In the model we employed, the variation in the vertical scale height across the entire range of Galactocentric distances covered by the data is described with a single number ($R_{\mathrm{flare}}^{-1}$), which makes it difficult to distinguish between specific scenarios. \citet{Lian2022MNRAS.513.4130L} remarked that using a model in which the scale height is a linear function of $R$ (rather than exponential) yields negligible flaring, and they proposed that a piece-wise model could highlight that flaring mostly affects the outer disc (beyond 10\,kpc). Using this approach (but not resolving the Milky Way into mono-abundance or mono-age populations), \citet{Mateu2018MNRAS.479..211M} proposed that flaring sets on beyond 11\,kpc. Numerical simulations have shown that flaring can be caused by multiple mechanisms \citep[see e.g.][and references therein]{GarciaDeLaCruz2021MNRAS.501.5105G}, such as external perturbations from satellite galaxies, stellar migration, or a tilt in the mass distribution of the Galactic halo \citep{Han2023NatAs.tmp..193H}. \citet{SotilloRamos2023MNRAS.523.3915S} reported ``a large variety of flaring flavours and amounts'' in the TNG50 simulations \citep[][]{Pillepich2019MNRAS.490.3196P}, which indicates that the flaring of the Milky Way disc might be difficult to capture with a simple, smooth density model.

We show in Sect.~\ref{sec:gspspec} that the \Gaia GSP-Spec abundances cannot distinguish between the high-$\alpha$ and low-$\alpha$ disc components. However, the strong flaring observed in the metal-rich population is probably not due to contamination by (geometrically thicker) $\alpha$-rich stars, which are rare in the outer disc due to their short scale length \citep[][]{Bovy16population}.     
At high metallicity ([M/H]>0.2\,dex), it is in fact difficult even for the APOGEE Mg-informed $\alpha$ abundances to distinguish the two disc components \citep[e.g.][]{Leung23}. The evolution of the flaring scale length with metallicity might then be investigated by simultaneously fitting two density models, providing a morphological definition of the $\alpha$-rich and $\alpha$-poor component of the disc at that metallicity \citep[resembling the traditional thin/thick distinction, see e.g.][]{Gilmore83,Juric08}. A clearer separation of the discs might be obtained with a sample using \Gaia XP abundances \citep{Andrae23} or the upcoming ground-based surveys WEAVE and 4MOST. This task is beyond the scope of this paper, however, which focuses on defining a selection function for the combination of two independent astronomical surveys. The $\alpha$-rich contaminants could also be separated through a kinematic selection \citep[e.g.][]{Bensby03}, whose selection function would have to be constructed as well.

\subsection{Uncertainty on the selection function}

We did not consider the uncertainty of the selection function itself. 
A possible extension of this work could involve parametrising the selection function and simultaneously fitting for the selection function and the density distribution in a hierarchical fit. The model could include a correlation between metallicity and absolute magnitude of the RC, with priors describing some assumptions on the age-metallicity relation across the Galaxy (effectively acting as a nuisance parameter).
For a more rigorous and quantitative comparison of the model uncertainties, we should also take the impact of the choice of extinction map into account: While we used the all-sky \citet{Lallement22} map, \citet{Bovy16population} and \citet{Mackereth17} used a combination of maps and did not cover the entire sky. We therefore stress that our reported best-fit values and corresponding uncertainties are ultimately tied to a given choice of extinction model.

\section{Conclusion} \label{sec:conclusion}

We have presented the construction of a selection function for a sample of \Gaia DR3 red clump stars with precise chemical abundances, for which we included a 3D extinction. We proposed a simple approximation to account for the metallicity dependence of the selection function. We provided a framework to define the combined selection function of \Gaia and APOGEE red clump stars that accounts for the different abundance scales of the two surveys.
The $\alpha$ abundances obtained from \Gaia spectra are not sufficient to clearly separate the $\alpha$-rich disc, and therefore, we focused on the $\alpha$-poor disc.

The APOGEE survey provides a sparser but much deeper coverage of the disc, while \Gaia brings better constraints on the vertical structure due to its dense all-sky coverage. We performed a model fitting using them separately abd then using both simultaneously to account for their respective selection functions.

We find a good overall agreement between the profiles traced by both surveys, and with previous studies. We also noted interesting differences stemming from the different spatial coverage of the surveys, notably some difference between the APOGEE and \Gaia results in the solar neighbourhood. The explanation for this may be that the density model we employed was too smooth to simultaneously account for the details of the density distribution on a small and large scale.

The \Gaia data allowed us to study the density profile of $\alpha$-poor mono-abundance populations within 2\,kpc. We confirmed the steepening of the density gradient with metallicity and the continuous decrease in scale height reported in studies based on APOGEE data. 

We make the APOGEE selection function used in this study available through the \texttt{GaiaUnlimited} Python package\footnote{\url{https://github.com/gaia-unlimited}}. The procedure we employed to construct the selection function of the \Gaia subsample of red clump stars is also explained in detail in a Jupyter notebook\footnote{\url{https://github.com/gaia-unlimited/gaiaunlimited/blob/main/docs/notebooks/RedClump_GSP-Spec_SF.ipynb}}, which contains the code that produced Figs.~\ref{fig:completenes_rc-gsp-spec}, \ref{fig:gaia_chem_SF_allsky_fourpanels}, \ref{fig:gaia_chem_SF_XY}, and \ref{fig:sf_with_metallicity}.






\section*{Acknowledgments}

TCG thanks David Hogg and Adrian Price-Whelan for their feedback on this manuscript.

This work has made use of data from the European Space Agency (ESA) mission {\it Gaia} (\url{https://www.cosmos.esa.int/gaia}), processed by the {\it Gaia} Data Processing and Analysis Consortium (DPAC, \url{https://www.cosmos.esa.int/web/gaia/dpac/consortium}). Funding for the DPAC has been provided by national institutions, particularly those participating in the {\it Gaia} Multilateral Agreement.

This work results from the GaiaUnlimited project, which has received funding from the European Union's Horizon 2020 research and innovation program under grant agreement No 101004110. The GaiaUnlimited project was started at the 2019 Santa Barbara Gaia Sprint, hosted by the Kavli Institute for Theoretical Physics at the University of California, Santa Barbara.
A.~R.~C. is partly supported by the Australian Research Council through a Discovery Early Career Researcher Award (DE190100656) and Discovery Project DP210100018.

 This work made use of the \href{https://www.overleaf.com/}{Overleaf platform} and the Python packages \texttt{astropy} \citep{2018AJ....156..123A}, \texttt{scipy} \citep{2020SciPy-NMeth}, \texttt{numpy} \citep{harris2020array}, \texttt{healpy}\footnote{\url{http://healpix.sourceforge.net}} \citep{2005ApJ...622..759G,Zonca2019}, and \texttt{matplotlib} \citep{Hunter:2007}. TCG acknowledges an extensive use of TOPCAT \citep{Taylor05} and Jupyter notebooks \citep{Kluyver2016jupyter}.

\bibliographystyle{aa}
\bibliography{refs}

\begin{appendix}

\section{Selection on renormalised unit weight error (\texttt{ruwe}) }

\label{app:ruwe}

\citet{Lindegren21astrosolution} introduced the renormalised unit weight error (\texttt{ruwe}) as a diagnostic to assess the reliability of the astrometric solution for individual sources. A maximum value of $1.4$ was later proposed as a simple filter to reject ill-behaved sources and inconsistent astrometric fits caused by unresolved binaries. 
Rejecting sources with \texttt{ruwe}$>$1.4 has become a common first step when selecting data from the \Gaia archive 
\citep[e.g.][]{Fabricius21, MaizApellaniz21, GaiaDR3chemcartoPVP, CruzReyes22, Medan23, Quintana23, Qin23, Alfonso23, Lee23, Ratzenboek23, Gontcharov23, Liao23}, but the selection biases introduced by this filtering have not been investigated to our knowledge.

\begin{figure*}
\includegraphics[width=0.99\textwidth]{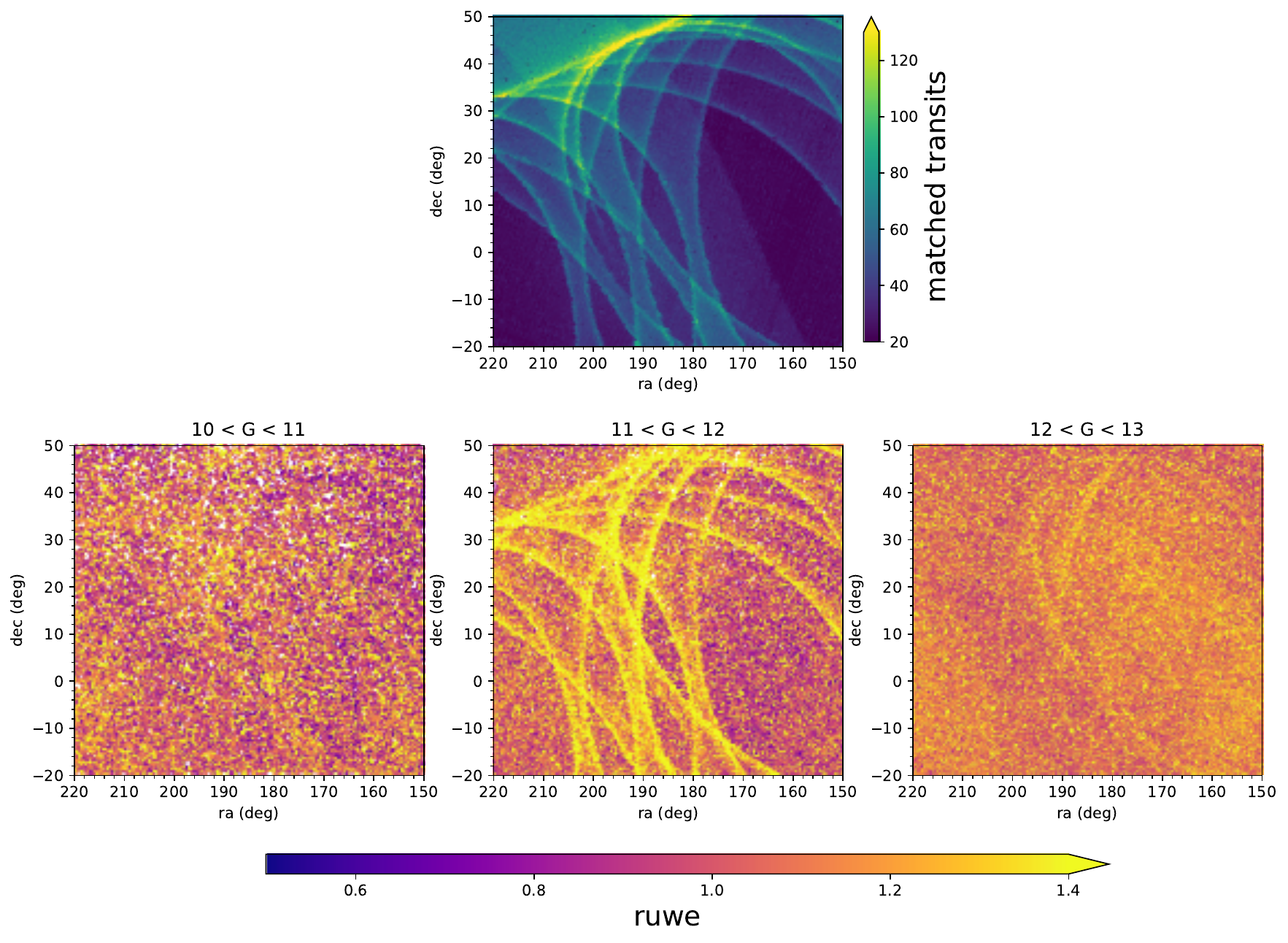} 
\caption{  Effect of magnitude and number of matched transits on \texttt{ruwe}. Top: Number of matched transits for sources with $G$<13 over a large patch of sky, showing the imprint of the \Gaia scanning law. Bottom: \texttt{ruwe} value for sources in this region in three different magnitude ranges. In areas with more than $\sim$120 \Gaia visits, removing sources with \texttt{ruwe} > 1.4 effectively discards almost all sources with $11 < G < 12$. \label{fig:ruwe_maps}}
\end{figure*}

\begin{figure}
\includegraphics[width=.99\columnwidth]{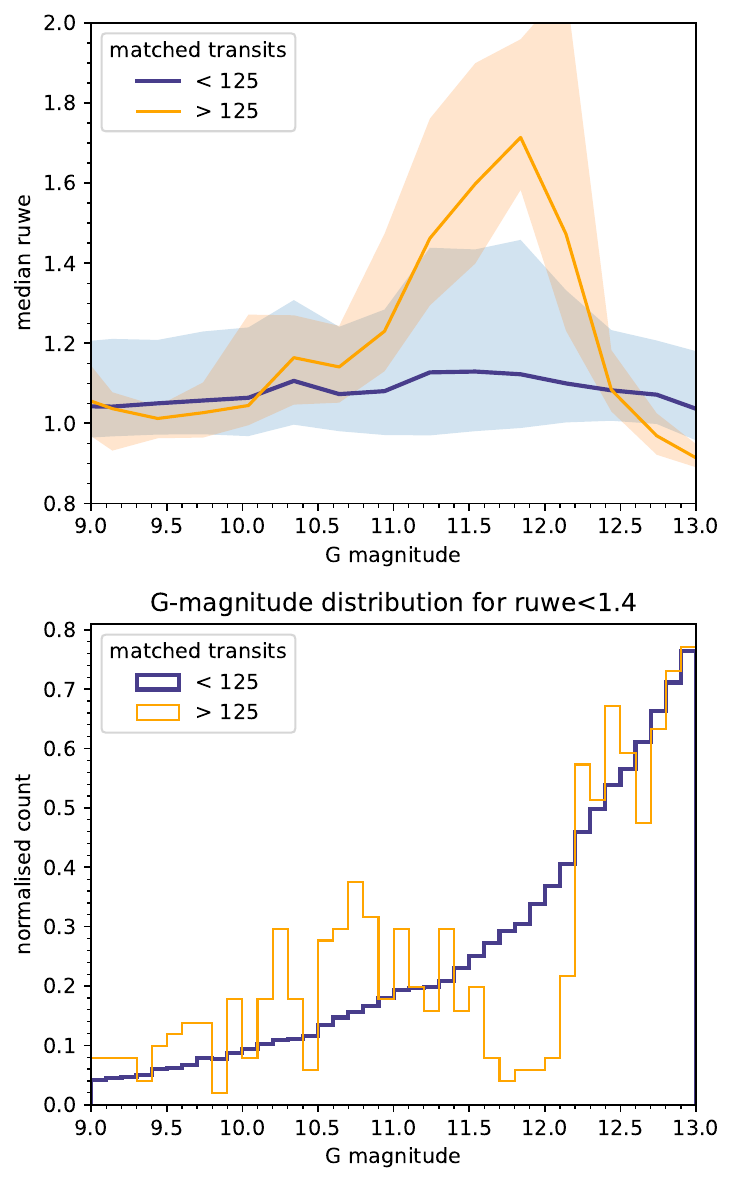} 
\caption{ Effect of magnitude and number of matched transits on \texttt{ruwe}. Top: Median \texttt{ruwe} value in bins of 0.3\,mag for sources with fewer and with more than 125 matched transits. The shaded area corresponds to the 25th to 75th percentile range. Bottom: $G$-magnitude distribution of the sources with \texttt{ruwe} $< 1.4$, and with fewer or more than 125 matched transits.   \label{fig:ruwe_plots}}
\end{figure}

The quantity \texttt{ruwe} was constructed so that its typical value was 1 across the whole sky and at all magnitudes.
In regions of the sky that received many \Gaia visits (usually more than 120), sources in the magnitude range $G$$\sim$11 to 12.5 have significantly higher \texttt{ruwe} values (Fig.~\ref{fig:ruwe_maps}), and removing sources with values above 1.4 creates an artificial dearth of sources, especially in the range $G$=11.5 to 12 (Fig.~\ref{fig:ruwe_plots}). We note that \citet{Golovin2023A&A...670A..19G} reported a very similar effect over the same magnitude range (shown in their Fig.~A.3) in their catalogue of nearby stars ($\varpi$>40\,mas).

The reason for this quirk happening within a very narrow magnitude range is currently not understood, but understanding the origin of this effect is not necessary in order to account for it. We note that it coincides with the magnitude limit ($G$=11.5) between the window classes 0A and 0B of the \Gaia video-processing units, and it might be related to calibration issues across this boundary. The \Gaia technical note \texttt{GAIA-C3-TN-LU-LL-124-01} describes how \texttt{ruwe} is computed from the unit weight error \texttt{uwe}, which is itself obtained from non-linear relations to various indicators of astrometric quality. The calculation of \texttt{ruwe} accounts for the fact that the residuals behave differently in different regimes of colour and magnitude, but it does not account for the number of scans, which might introduce this glitch for sources with a unfortunate combination of $G$ magnitude and transit numbers.

\section{Profile fit using only the most complete bins}  \label{app:80pct}

\begin{figure*}
\includegraphics[width=0.99\textwidth]{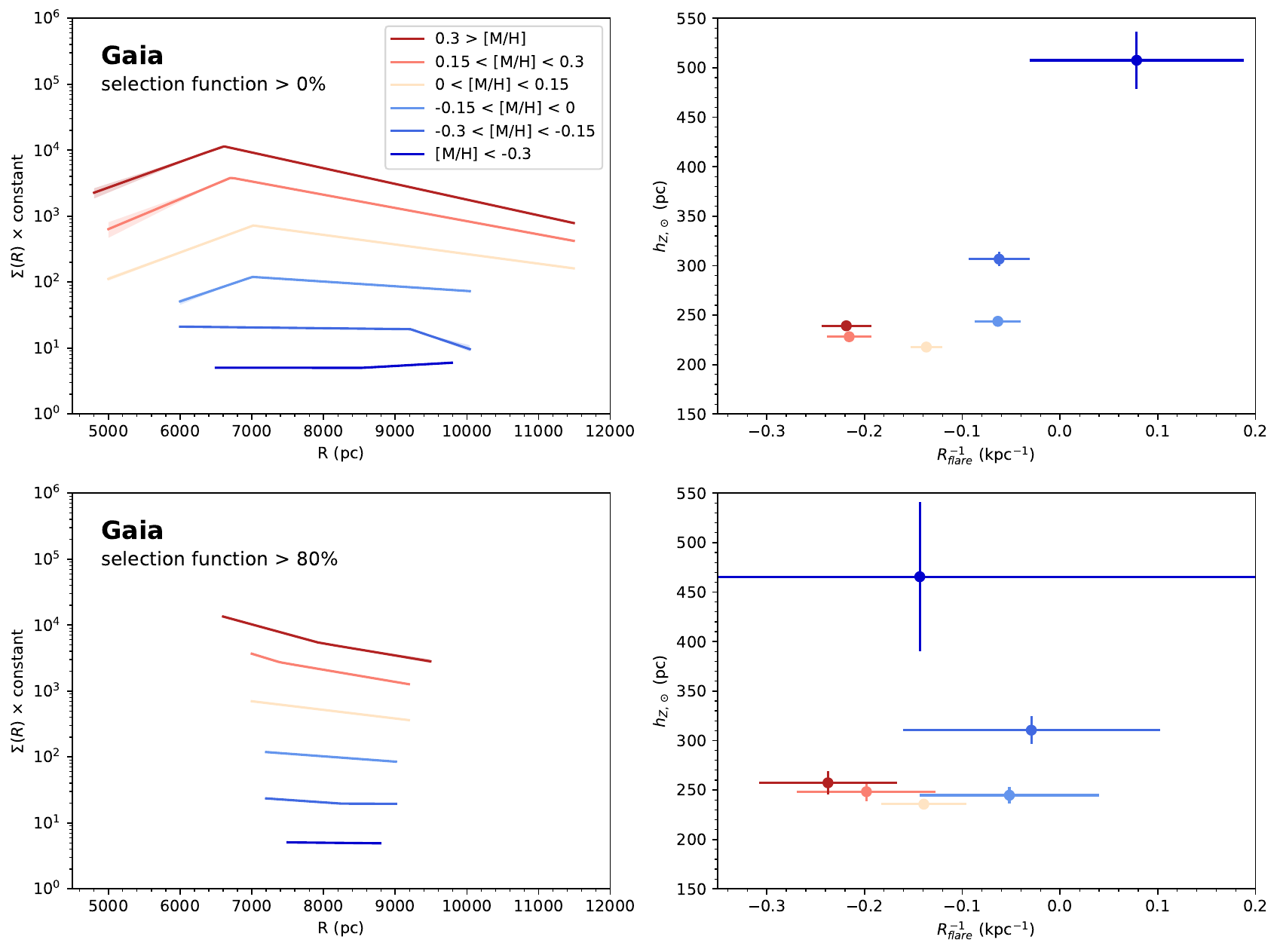} 
\caption{   \label{fig:density_gaia_80} Density profile for RC-GSP-Spec stars built upon any (top) and high completeness levels (80\%; bottom). The top row is identical to the middle row of Fig.~\ref{fig:density_profiles_all}. The high-completeness requirement greatly reduces the coverage in Galactocentric radius $R$, but the profiles remain statistically similar.
}
\end{figure*}

In Sect.~\ref{sec:SF_met} we commented on the metallicity dependence of the RC-GSP-Spec selection function and proposed a simple approximation to account for this dependence. The $G$-magnitude distributions shown in the bottom panel of Fig.~\ref{fig:sf_with_metallicity} suggest that the drop in the selection function from $\sim$80\% to 0\% completeness is steeper at a given metallicity than when averaged over the entire sample. In order to sidestep this uncertainty on the completeness profile of individual metallicity ranges, we attempted to only perform the fitting of the density profile over the volume of space where each sample is considered to be over 80\% complete. The results are shown in Fig.~\ref{fig:density_gaia_80}.

This requirement greatly reduced the coverage in Galactocentric radius $R$, especially for the low-metallicity sample. The density profiles are shown in the bottom left panel of Fig.~\ref{fig:density_gaia_80}. They essentially match the trends shown in Fig.~\ref{fig:gaia_density_corrected}, where we see a flat (at low metallicity) or monotonic increase towards the Galactic centre (at high metallicity) density in the solar neighbourhood. 

The reduced coverage in $R$ leads to very large uncertainties on the flaring scale length $R_{flare^{-1}}$. We remark that because the distance range covered by this restricted data is still several times larger than the vertical scale height, we are still able to accurately recover the values of $h_{Z,\odot}$$\sim$450\,pc for the [M/H] < -0.3\,dex population, $\sim$300\,pc for -0.3 < [M/H] < -0.15, and $\sim$250\,pc for the more metal-rich stars.

%
\section{Vertical density distribution of the metal-rich population} \label{app:hZmetalrich}

In the analytic model presented in Sect.~\ref{sec:density_model}, the vertical density distribution $\zeta(Z|R)$ follows an exponential profile with scale length varying with Galactocentric distance $R$,

\begin{equation}
h_Z^{-1}(R) = h_{Z,\odot}^{-1} \exp{(R_{\mathrm{flare}}^{-1}[R-R_0])}.
\end{equation}

\begin{figure*}
\includegraphics[width=0.99\textwidth]{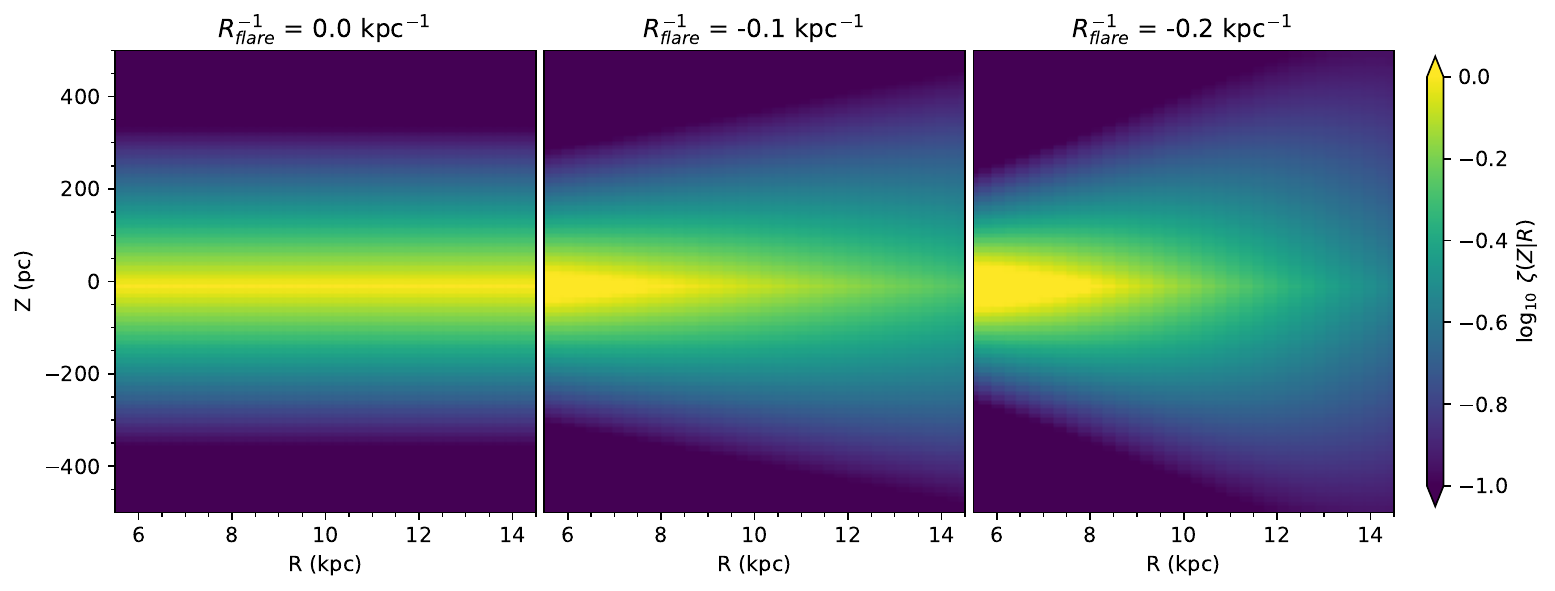} 
\caption{  Vertical density distribution $\zeta$ for a flared exponential disc with $h_{Z,\odot}$=150\,pc as a function of $R$ and $Z$ for three different flaring scale lengths. For the purpose of this plot, the densities in all three panels are normalised so that $\zeta(Z=0|R_{\odot})$=1. \label{fig:illustration_Rflare}
}
\end{figure*}

\begin{figure*}
\includegraphics[width=0.99\textwidth]{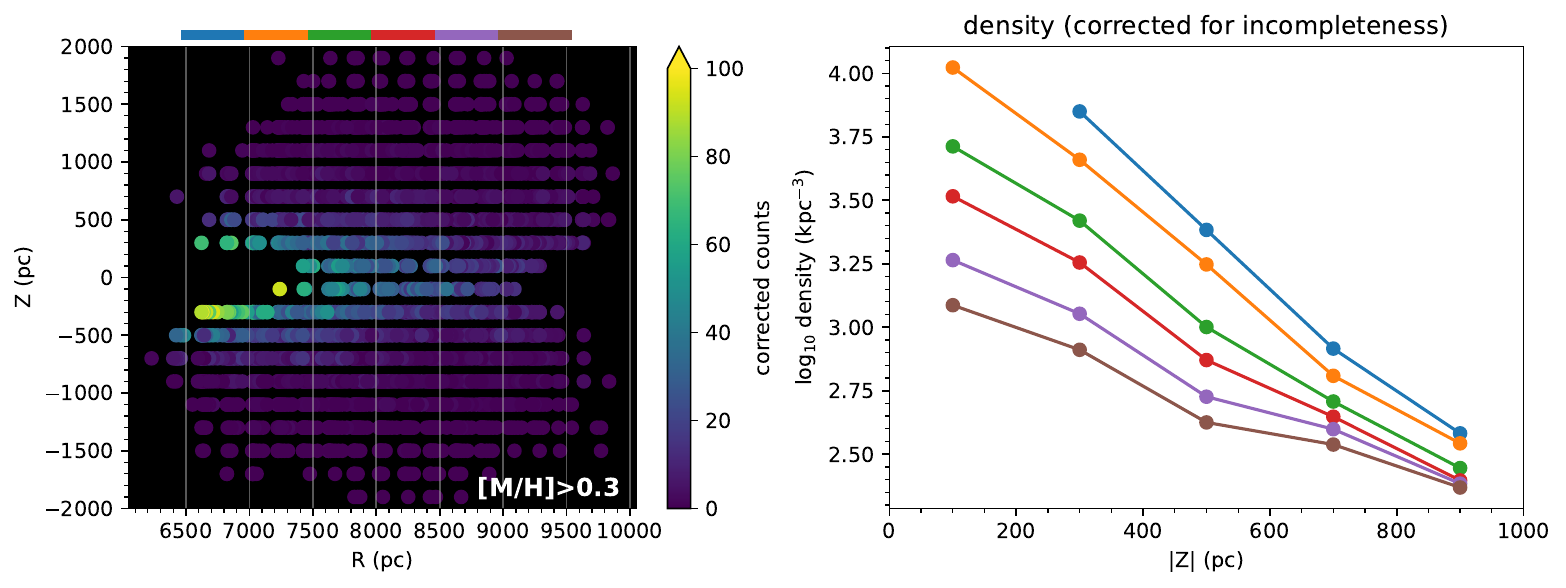} 
\caption{  Left: Number of observed RC stars with [M/H]>0.3\,dex computed in $(X,Y,Z)$ bins of 200\,pc, corrected for incompleteness (where the completeness is above 70\% only), and displayed as a function of $R$ and $Z$. The Sun is at $R$=8.15\,kpc. Right: Log-density as a function of distance from the Galactic mid-plane $|Z|$ computed in six ranges of $R$ (colour-coded as shown above the left panel). The flattening of the density profile with increasing $R$ is interpreted by our model as a sign of flaring. We reiterate that the observed counts should only be corrected for completeness for data exploration, and the model fitting procedure of Sect.~\ref{sec:density_model} is instead performed by applying the selection function to the model.
\label{fig:flaring_metal_rich}}
\end{figure*}

With this definition, a negative value of $R_{\mathrm{flare}}^{-1}$ causes $h_Z^{-1}$ to decrease with $R$, causing the scale height $h_Z$ to increase. This effect is illustrated in Fig.~\ref{fig:illustration_Rflare} for three chosen values of $R_{\mathrm{flare}}^{-1}$. The density profiles we obtain from \Gaia data (Fig.~\ref{fig:density_profiles_all}) show a strong degree of flaring, with the most metal-rich populations reaching, $R_{\mathrm{flare}}^{-1}$$\sim$0.2\,kpc$^{-1}$. This strong flaring is illustrated (qualitatively) in Fig.~\ref{fig:flaring_metal_rich}, where the right panel shows that the vertical density drop is steeper in the inner than in the outer disc.

Using APOGEE DR12 data, \citet{Bovy16population} found that the $\alpha$-poor disc only exhibits moderate flaring with $R_{\mathrm{flare}}^{-1}$=-0.1 at all metallicities, but they pointed out that the footprint of the APOGEE survey is dominated by low-latitude fields that provide little leverage on the vertical density profile. 
The dense coverage offered by \Gaia in the solar neighbourhood provides much stronger constraints.

\section{Best-fit parameter table}

\begin{table*}
\begin{center}
	\caption{ \label{tab:mcmc} Best-fit parameters for the Milky Way density profile.}

\begin{tabular}{ c c c c c c}
\hline
metallicity & $h_{R,\text{in}}^{-1}$ & $h_{R,\text{out}}^{-1}$ & $R_{\text{peak}}$ & $h_{Z,\odot}$  & $R_{\mathrm{flare}}^{-1}$  \\
\hline
\hline
\multicolumn{6}{c}{APOGEE only} \\
\hline
[M/H] < -0.3 & -0.47 $\pm$ 0.11 & 0.26 $\pm$ 0.22 & 10.24 $\pm$ 0.50 & 340 $\pm$ 68 & -0.10 $\pm$ 0.07   \\
-0.3 < [M/H] < -0.15 & -0.28 $\pm$ 0.12 & 0.60 $\pm$ 0.15 & 9.96 $\pm$ 0.25 & 257 $\pm$ 36 & -0.10 $\pm$ 0.06   \\
-0.15 < [M/H] < 0 & -0.09 $\pm$ 0.21 & 1.01 $\pm$ 0.13 & 9.71 $\pm$ 0.40 & 231 $\pm$ 26 & -0.11 $\pm$ 0.05   \\
0 < [M/H] < 0.15 & -0.06 $\pm$ 0.17 & 1.01 $\pm$ 0.26 & 8.16 $\pm$ 0.67 & 214 $\pm$ 24 & -0.15 $\pm$ 0.06   \\
0.15 < [M/H] < 3 & -0.49 $\pm$ 0.14 & 1.15 $\pm$ 0.41 & 6.50 $\pm$ 0.28 & 208 $\pm$ 29 & -0.17 $\pm$ 0.06   \\

[M/H] > 0.3 & 0.07 $\pm$ 0.27 & 1.14 $\pm$ 0.83 & 6.17 $\pm$ 0.96 & 236 $\pm$ 78 & -0.20 $\pm$ 0.15   \\
\hline
\multicolumn{6}{c}{\textit{Gaia} only} \\
\hline
[M/H] < -0.3 & 0.01 $\pm$ 0.19 & -0.32 $\pm$ 0.08 & 8.50 $\pm$ 0.26 & 507 $\pm$ 29 & 0.08 $\pm$ 0.11   \\
-0.3 < [M/H] < -0.15 & 0.06 $\pm$ 0.78 & 1.94 $\pm$ 0.03 & 9.21 $\pm$ 0.11 & 306 $\pm$ 7 & -0.06 $\pm$ 0.03   \\
-0.15 < [M/H] < 0 & -1.96 $\pm$ 0.02 & 0.38 $\pm$ 0.56 & 7.01 $\pm$ 0.05 & 243 $\pm$ 3 & -0.06 $\pm$ 0.02   \\
0 < [M/H] < 0.15 & -2.12 $\pm$ 0.02 & 0.77 $\pm$ 0.19 & 7.02 $\pm$ 0.02 & 217 $\pm$ 2 & -0.14 $\pm$ 0.02   \\
0.15 < [M/H] < 3 & -2.42 $\pm$ 0.04 & 1.06 $\pm$ 0.86 & 6.71 $\pm$ 0.07 & 228 $\pm$ 4 & -0.22 $\pm$ 0.02   \\

[M/H] > 0.3 & -2.05 $\pm$ 0.03 & 1.26 $\pm$ 0.54 & 6.61 $\pm$ 0.06 & 239 $\pm$ 5 & -0.22 $\pm$ 0.03   \\
\hline
\multicolumn{6}{c}{\Gaia + APOGEE} \\
\hline
[M/H] < -0.3 & -0.20 $\pm$ 0.04 & 0.58 $\pm$ 0.05 & 8.73 $\pm$ 0.11 & 482 $\pm$ 20 & -0.05 $\pm$ 0.03   \\
-0.3 < [M/H] < -0.15 & -0.06 $\pm$ 0.05 & 0.80 $\pm$ 0.04 & 8.71 $\pm$ 0.08 & 307 $\pm$ 6 & -0.08 $\pm$ 0.02   \\
-0.15 < [M/H] < 0 & -1.38 $\pm$ 0.02 & 0.63 $\pm$ 0.29 & 7.18 $\pm$ 0.06 & 244 $\pm$ 3 & -0.11 $\pm$ 0.02   \\
0 < [M/H] < 0.15 & -1.66 $\pm$ 0.02 & 0.85 $\pm$ 0.25 & 7.07 $\pm$ 0.05 & 220 $\pm$ 2 & -0.16 $\pm$ 0.02   \\
0.15 < [M/H] < 3 & -1.12 $\pm$ 0.02 & 1.11 $\pm$ 0.17 & 6.83 $\pm$ 0.01 & 229 $\pm$ 4 & -0.23 $\pm$ 0.02   \\

[M/H] > 0.3 & -1.52 $\pm$ 0.03 & 1.28 $\pm$ 0.27 & 6.66 $\pm$ 0.03 & 240 $\pm$ 6 & -0.22 $\pm$ 0.02   \\
\hline
\end{tabular}
\tablefoot{ The parameters correspond to the models shown in Fig.~\ref{fig:density_profiles_all}, whose analytic form is given in Sect.~\ref{sec:density_model}. The uncertainties are the 95\% confidence intervals from MCMC sampling.}
\end{center}
\end{table*}

\end{appendix}
\end{document}